\RequirePackage{fix-cm}
\documentclass[twocolumn,epjc3]{svjour3}  
\pdfoutput=1
\RequirePackage{graphicx}
\usepackage{amsmath}
\usepackage{amssymb}
\usepackage{array}
\usepackage{calc}
\usepackage{longtable}
\usepackage{multirow}
\usepackage{wasysym}
\usepackage{xspace}
\usepackage{todonotes}
\usepackage{placeins}
\usepackage[pdfborder={0 0 0}]{hyperref}
\usepackage[format=hang,labelfont=bf,hypcap=true]{caption}
\usepackage{subfig}
\usepackage{mathtools}
\usepackage{lmodern}
\usepackage{mciteplus}

\newcommand{\RM}[1]{\MakeUppercase{\romannumeral #1{}}}

\newcommand{\gev}{\ensuremath{\,\mathrm{GeV}}}
\newcommand{\tev}{\ensuremath{\,\mathrm{TeV}}}
\newcommand{\DELTA}{\ensuremath{\mathrm{\Delta}}}

\journalname{Eur. Phys. J. C}
\DeclarePairedDelimiter\abs{\lvert}{\rvert}

\begin{document}

\title{NLO QCD predictions for $Z+\gamma$ + jets production with Sherpa}
\author{Johannes Krause\thanksref{e1,addr1} \and Frank Siegert\thanksref{e2,addr1}}
\thankstext{e1}{e-mail: johannes.krause1@tu-dresden.de}
\thankstext{e2}{e-mail: frank.siegert@cern.ch}
\institute{Institut f{\"u}r Kern- und Teilchenphysik, TU Dresden, D--01062 Dresden, Germany\label{addr1}}
\date{Received: date / Accepted: date}

\maketitle



\begin{abstract}

  We present precise predictions for prompt photon production in association
  with a $Z$ boson and jets. They are obtained within the Sherpa framework
  as a consistently merged inclusive sample. Leptonic decays of the $Z$ boson
  are fully included in the calculation with all offshell effects. Virtual
  matrix elements are provided by OpenLoops and parton shower effects are
  simulated with a dipole parton shower.
  Thanks to the NLO QCD corrections included not only for inclusive $Z\gamma$
  production but also for the $Z\gamma$ + 1-jet process we find significantly
  reduced systematic uncertainties and very good agreement with experimental
  measurements at $\sqrt{s}=8\tev$. Predictions at $\sqrt{s}=13\tev$ are
  displayed including a study of theoretical uncertainties.
  In view of an application of these simulations within LHC experiments, we
  discuss in detail the necessary combination with a simulation of the
  $Z$ + jets final state. In addition to a corresponding prescription we
  introduce recommended cross checks to avoid common pitfalls during the
  overlap removal between the two samples.
  
\end{abstract}

\pagebreak
\setcounter{tocdepth}{3}
\tableofcontents

\section{Introduction}

The production of a $Z$ boson is one of the standard candle
processes at hadron colliders like the LHC. The massive boson is often
produced in association with photons, which are typically low-energetic
or collinear with charged final state particles. Cases where one
of the photons happens to be well-isolated and high-energetic can be
regarded as an individual important final state, $Z\gamma$ production.

$Z\gamma$ production plays an important role both as a signal and as a
background process at the LHC. The absence of couplings of the photon to
the uncharged $Z$ boson in the Standard Model can be probed by measurements
in this channel, resulting in differential cross sections and limits on
anomalous couplings from
LEP experiments~\cite{Achard:2004ds,*Abdallah:2007ae,*Abbiendi:2000cu,*Abbiendi:2004bf},
Tevatron experiments~\cite{Abazov:2009cj,*Abazov:2011qp,*Aaltonen:2011zc},
and by \mbox{ATLAS}~\cite{Aad:2013izg,*Aad:2016sau,*Aaboud:2017pds} and
CMS~\cite{Chatrchyan:2013nda,*Chatrchyan:2013fya,*Khachatryan:2015kea,*Khachatryan:2016yro}.

The $Z\gamma$ process also constitutes an irreducible background in the search
for the Higgs boson~\cite{Djouadi:1996yq} or new massive gauge bosons decaying
to $Z\gamma$~\cite{Aaboud:2016trl,*Khachatryan:2016odk} or in more inclusive searches in final states containing a
photon and missing transverse momentum~\cite{Aaboud:2016uro,*Aaboud:2017dor,Khachatryan:2016ojf,*Sirunyan:2017ewk}.

Theoretical predictions for $Z\gamma$ production can be divided into onshell and
offshell calculations.
Results for {\it onshell} $Z\gamma$ production leave out the decays of the
$Z$\ boson or include them only in a narrow-width or pole approximation.
Beyond the leading-order results~\cite{Renard:1981es},
the first onshell higher-order calculations included
NLO QCD~\cite{Ohnemus:1992jn,*Ohnemus:1994qp,*Baur:1997kz}
and, more recently, also NNLO QCD corrections~\cite{Grazzini:2013bna}.
In reality, the $Z$\ boson is unstable and is thus never produced as an onshell
final-state particle. Recent calculations take this finite width
into account and provide predictions for the {\it offshell}\, $\ell\ell\gamma$
final state. The most accurate offshell predictions contain
NNLO QCD~\cite{Grazzini:2015nwa,*Campbell:2017aul} and NLO EW corrections~\cite{Denner:2015fca}.

Experimental analyses at the LHC rely heavily on theoretical predictions
for their signal and background processes. Fixed-order predictions as listed
above can provide such an input only to some extent. While they describe the
dominant features of the given final state objects at the parton level, they
are not made to simulate a realistic behaviour of the hadronic final state.
Monte-Carlo event generator programs on the other hand combine fixed-order predictions and an
approximate all-order resummation of QCD corrections to enable a full simulation
at the hadron level. Different approaches and programs are available, but the
simulation currently in use in LHC experiments for $Z\gamma$ production is generated with
leading-order multi-leg generators using approaches like CKKW(-L) or MLM
merging~\cite{Catani:2001cc,*Lonnblad:2001iq,*Krauss:2002up,*Mangano:2001xp,
  *Alwall:2007fs,*Hamilton:2009ne,*Hoeche:2009rj}. For the related process
of $W\gamma$ production, an implementation of a NLO QCD calculation of the
inclusive process matched to a parton shower exists within the Powheg
framework~\cite{Barze:2014zba}.

This article applies a NLO accurate multi-leg merging formalism to
the processes of $Z\gamma$ and $Z\gamma$+jet
production\footnote{Even though the process is denoted with the shorthand
  $Z\gamma$, the calculations throughout this paper include the full offshell
  $\ell\ell\gamma$ final state.}
for the first time.
After a review of the relevant methods in Section~\ref{sec:methods} we present
our computational setup and results comparing LO and NLO accurate merged
predictions to each other and to experimental data in
Section~\ref{sec:results_mepsnlo}. A special aspect relevant for the application
of multi-jet merged $Z\gamma$ samples in experimental analyses is a combination
of $Z\gamma$+jets and $Z$+jets. Recommended techniques and cross checks for
such a combination are implemented and discussed in
Section~\ref{sec:results_fsr}.

\section{Methods}
\label{sec:methods}

\subsection{Matching and merging with \textsc{Sherpa} at NLO}
\label{subsec:matching_merging}

To obtain NLO accurate multi-jet-merged predictions the ``MEPS@NLO''
formalism~\cite{Hoeche:2012yf} is applied to $Z\gamma$ production within the \textsc{Sherpa}
framework. It combines two essential ingredients, NLO + parton-shower matching
and multi-jet merging, which are briefly summarised in the following.

For the combination of NLO-accurate matrix-element calculations with a parton
shower (PS), a matching procedure to avoid double counting
of the QCD emission effects at $\mathcal{O}(\alpha_s)$ is needed. Here, the prescription
proposed in~\cite{Hoeche:2011fd,*Hoeche:2012ft} is applied, both to
$pp\to Z\gamma$ and $pp\to Z\gamma$+jet simulations.
It is based on the original MC@NLO method~\cite{Frixione:2002ik} but extends
it to a fully colour-correct formulation also in the limit of soft emissions.

Both simulations and further PS emissions are then consistently merged into
one inclusive $pp\to Z\gamma$+0,1j@NLO sample using the
``MEPS@NLO'' me\-thod~\cite{Hoeche:2012yf}. Shower emissions above a pre-defined
separation criterion, $Q_\textrm{CUT}$, are vetoed in the lower multiplicity
contribution, and Sudakov factors are applied where appropriate such as to
make this contribution exclusive and allow the combination with a higher jet
multiplicity. This applies not only to the emissions from the parton
shower but to all contributions of the NLO-matched emission, including the hard
remainder (``$\mathbb{H}$ events''). In the application of the Sudakov factors
care has to be taken to remove the $\mathcal{O}(\alpha_s)$ contribution which
is already present in the NLO(-matched) emission.

Beyond the processes simulated at NLO accuracy, higher-multiplicity processes
can be added to the simulation at LO accuracy to improve the modelling of
high jet multiplicities beyond the parton shower approximation.

Similar to merging methods at LO, an appropriate scale choice for the
evaluation of multi-jet configurations is obtained by statistically
identifying a parton shower history in the matrix-element final state.
To that end, the parton shower is run in reverse mode, i.e.\ the closest parton
pair is identified according to the shower splitting probabilities and then
recombined using the kinematical properties of the shower.
When applied recursively, this clustering results in a \emph{core process},
e.g.\ $pp\to Z\gamma$, and in an ordered history of shower emissions. The
factorisation scale is then determined by a typical momentum transfer within
the core process (core scale), and the renormalisation scale is calculated from
the $n$ identified branchings to resemble
$\alpha_s^n(\mu_R)=\prod_{i=1}^n\alpha_s(k_{\perp,i})$,
where $k_{\perp,i}$ is a suitably scaled relative transverse momentum encountered in the
$i$th splitting. For events with very hard QCD emissions, the clustering may
include electroweak combinations to preserve the ordering in the history
(``\emph{inclusive clustering}''). In such events, the core process can also
contain jets and contribute to the renormalisation scale accordingly.

\subsection{Soft photon resummation with YFS}
\label{subsec:yfs}

\newcommand{\cnv}{\ensuremath{n_\mathrm{V}}}
\newcommand{\cnr}{\ensuremath{n_\mathrm{R}}}
\newcommand{\cng}{\ensuremath{n_{\gamma}}}

The YFS-algorithm \cite{Yennie:1961ad} 
describes a possibility to resum soft logarithmically enhanced photon radiation to all orders in a process-independent manner. 
This algorithm is implemented in \textsc{Sherpa} for both, leptonic decays of $W$ / $Z$ bosons and hadron decays. 
Details of this implementation are given in~\cite{Schonherr:2008av} and briefly summarised here. 
The decay width for a decay of a particle $i$ with mass $M$ into a final state $f$, 
corrected for the radiation of  \cnr\  real and \cnv\ virtual photons, reads

\begin{equation}
\begin{split}
2M \Gamma &= \sum \limits_{\cnr=0}^{\infty} \int \mathrm{d} \tilde{\Phi} \left| \sum \limits_{\cnv}^{\infty}  \mathcal{M}_{\cnr}^{\cnv + \frac{1}{2}\cnr}  \right|^2. 
\end{split}
\label{eq:dwfull}
\end{equation}

Here, $\mathcal{M}_{\cnr}^{\cnv + \tfrac{1}{2}\cnr}$ describes a decay matrix element with additional \cnr\ real and \cnv\ virtual photons and  $\tilde{\Phi}$ denotes the corresponding phase space.  
As shown in \cite{Yennie:1961ad}, the soft limits of all these (virtual and real) matrix elements can be resummed and factorized out. 
The corresponding infrared divergences cancel order by order and result in the finite YFS form factor $Y(\mathrm{\Omega})$.
$Y(\mathrm{\Omega})$ includes the soft limits of all virtual and real contributions. 
However, only the divergent part of the real contributions is retained and separated by the symbolic cut off $\mathrm{\Omega}$ from the non-divergent one. 
Eq.~\eqref{eq:dwfull} can then be approximated with 

\begin{equation}
2M\Gamma = \sum \limits_{\cng} \frac{1}{\cng} \int \mathrm{d} \Phi e^{Y(\mathrm{\Omega})} \prod \limits_{i=1}^{\cng} S(k_i) \Theta(k_i, \mathrm{\Omega}) \beta_0^0 \mathcal C.
\label{eq:dwyfs}  
\end{equation} 

The eikonals $S(k)$ include all contributions for the emission of a real photon with momentum $k$.
Since the divergent, real part is already included in $Y(\mathrm{\Omega})$, these eikonals will only be integrated in the non divergent phase space.
The number of additional, resolved photons is denoted with $\cng$, while $\beta_0^0$ is the undressed matrix element without any real or virtual photons and $\Phi$ the corresponding phase space.
The YFS algorithm can in principle be improved order by order with exact, process dependent real and virtual matrix elements. These possible corrections are incorporated in the factor $\mathcal C$, which is equal to one in case of no corrections.
In \textsc{Sherpa}, higher order corrections are available either in an approximative way using collinear splitting kernels or as exact matrix elements. The first option is independent of the process whereas the second one is limited to a few cases, including the decay of a vector boson into two fermions. 
 
It is worth noting that in a Monte Carlo code photons radiated by YFS are --  in contrast to e.g. QED showers -- unordered.

\subsection{Isolated photons}
\label{subsec:iso_photo}
Matching theoretical predictions with experimental measurements including isolated photons has turned out to be a non-trivial problem when going beyond leading order in QCD. It is no longer possible to isolate the photon completely from all kind of hadronic activity since this would constrain the phase space of soft gluon radiation and destroy the cancellation of infrared singularities. 
As a consequence, many experiments allow a small fraction of hadronic energy within the isolation cone. 
However, this relaxed cone criterion requires special attention when calculating theoretical cross sections. An observable constructed with such an isolation criterion includes a divergence when the photon gets collinear to a massless quark. This divergence is of QED origin and does not cancel within the pertubative QCD calculation. In principle there are two common ways to solve this problem, either by absorbing the divergences into fragmentation functions or by using a smooth isolation criterion. Such a smooth isolation criterion has been proposed in \cite{Frixione:1998jh} and is also used in this paper. This criterion suppresses the divergent collinear contribution by limiting the maximal transverse energy $E_\perp^{\mathrm{max}}$ close to the photon axis, 
\begin{equation}
 E_\perp^{\mathrm{max}} \leq  \epsilon  p_\perp^{\gamma} \left( \frac{1-\cos(r)}{1-\cos(R)} \right)^n .
\label{eq:frix}
\end{equation}
Here, $p_\perp^{\gamma}$ is the transverse momentum of the photon and $\epsilon$, $n$ and $R$ are parameters which define the final shape of the smooth cone. 
$E_\perp^{\mathrm{max}}$ is defined as the sum of the transverse energies of all partons present at matrix element level within a cone of radius $r$ around the photon axis.
The condition in Eq.~\eqref{eq:frix} has to be fulfilled for all cones with
\begin{equation}
 r=\sqrt{(\DELTA\eta)^2 + (\DELTA \phi)^2} < R
\end{equation}
and ensures that $E_\perp^{\mathrm{max}}$ converges smoothly to zero for $ r \rightarrow 0$.

\section{NLO-accurate multi-jet predictions for $Z\gamma$ production}
\label{sec:results_mepsnlo}

\subsection{Setup}
\label{subsec:setup}
All results in this publication are obtained with the Monte Carlo event generator \textsc{Sherpa}~2.2.2~\cite{Gleisberg:2008ta} using merged calculations. 
Two setups will be compared in the following:
\begin{description}
\item[MEPS@NLO]\hfill\\
  $pp \rightarrow e^+ e^- \gamma + 0,1\mathrm{jets@NLO} + 2,3\mathrm{jets@LO} $,
\item[MEPS@LO]\hfill\\
  $pp \rightarrow e^+ e^- \gamma + 0,1,2,3\mathrm{jets@LO}$.
\end{description}
Therein, jet refers to an additional parton in the matrix element and LO/NLO denotes the accuracy of the corresponding multiplicity. 

The matrix elements are calculated by the internal
matrix element generators \textsc{Amegic++}~\cite{Krauss:2001iv} and \textsc{Comix}~\cite{Gleisberg:2008fv}. Virtual diagrams are calculated by \mbox{\textsc{OpenLoops 1.3.1}}~\cite{Cascioli:2011va}, using  \mbox{\textsc{CutTools}}~\cite{Ossola:2007ax} and \mbox{\textsc{OneLoop}}~\cite{vanHameren:2010cp}.  
In all MEPS@NLO setups \textsc{Amegic++} is used only for Born-like processes. All real-subtracted
contributions and the leading order diagrams of higher multiplicity are calculated by
\textsc{Comix}.
The merging cut $Q_\mathrm{CUT}$ is set to 30\gev.

Where not explicitly stated otherwise, scales are determined by the \emph{inclusive}
clustering algorithm described in Sec.~\ref{subsec:matching_merging} (\texttt{STRICT\_METS}).
The core scale is calculated according to the core process as
\begin{equation}
\label{eq:def_core_scale}
  \mu^2_\text{core}=\begin{cases}
      m^2_{Z}            & \text{for $Z$},\\
      m^2_{Z\gamma}            & \text{for $Z\gamma$},\\
    \tfrac{1}{4}m^2_{\perp,Z/\gamma}             & \text{for $Z/\gamma$+jet},\\
    \tfrac{1}{4}\tfrac{-1}{1/\hat{s}+1/\hat{t}+1/\hat{u}}  & \text{for jet+jet},\\
    \tfrac{1}{4}\bigl(m_{\perp,Z\gamma} ~ +  & \\
    \hphantom{\tfrac{1}{4}\bigl(} \sum_{\text{jets}} m_{\perp,\text{jet}} \bigr)^2   & \text{for unordered $Z\gamma$+jets.}\\
  \end{cases}
\end{equation}
This corresponds to the default core scale implementation in \textsc{Sherpa}~2.2.

The electroweak couplings are evaluated using a mixed scheme as recommended in~\cite{Butterworth:2014efa}.
First, all couplings are calculated using the $G_\mu$ scheme.
In this scheme, the coupling constant is evaluated as function of the Fermi constant $G_\mu$ and the masses of $W$ and $Z$, 
\begin{equation}
\alpha_{G_\mu} = \frac{\sqrt{2}}{\pi} G_\mu M_W^2 \left(1 - \frac{M_W^2}{M_Z^2}\right). 
\end{equation}
This behaviour effectively resums contributions which arise when evolving the electroweak coupling to the electroweak scale and is a common choice for processes involving heavy $W$ or $Z$ bosons. 
However, in the $V\gamma$ processes an additional external -- i.e. on-shell -- photon is present. 
Taking this into account one electroweak coupling should be evaluated at $\alpha(0)$. This is achieved by a global reweighting with $k=\alpha(0) / \alpha_{G_\mu}$. 

Unstable particles are described using the complex mass scheme~\cite{DENNER200622}.
Following~\cite{Denner:2015fca},  on-shell masses and widths are converted to pole values and result in

\begin{equation}
\begin{split}
M_W=80.3580\,\gev, ~~~ \Gamma_W=2.0843\,\gev, \\
M_Z=91.1535\,\gev, ~~~ \Gamma_Z=2.4943\,\gev.
\end{split}
\end{equation} 
These values are used for all calculations.

As parton distribution functions NNPDF3.0~\cite{Ball:2014uwa} sets are used,
taking  the NLO set for MEPS@LO calculations and the NNLO set for MEPS@NLO. The
running of $\alpha_\mathrm{s}$ and its value at $M_\mathrm{Z}$ are thereby set
according to these PDF sets, resulting in
$\alpha_\mathrm{s}(M_\mathrm{Z})=0.118$ and a running at two(three) loop order when
using the NLO(NNLO) sets.
YFS is set active and includes matrix element corrections for further photon emissions.

For the event generation, all parton level cuts are set to be more inclusive
than the respective analysis cuts. As described in Section~\ref{subsec:iso_photo},
 it is not possible to use the experimental isolation
criterion, instead the smooth cone criterion is used and validated for two
different sets of parameters.

When comparing to experimental data the simulation is performed including the default multiple interactions~\cite{Sjostrand:1987su,*Alekhin:2005dx} and hadronisation models~\cite{Winter:2003tt}. The final state analyses are done within the \textsc{Rivet} framework~\cite{Buckley:2010ar}.
 

\subsection{Predictions for $\sqrt s=$13\tev}
\label{subsec:results_13_dir}

\subsubsection{Merging cut variation}
\begin{figure*}
  \subfloat[$k_\perp$ jet resolution $0 \rightarrow 1$ \label{subfig:qcut_a}]{\includegraphics[width=0.33\linewidth]{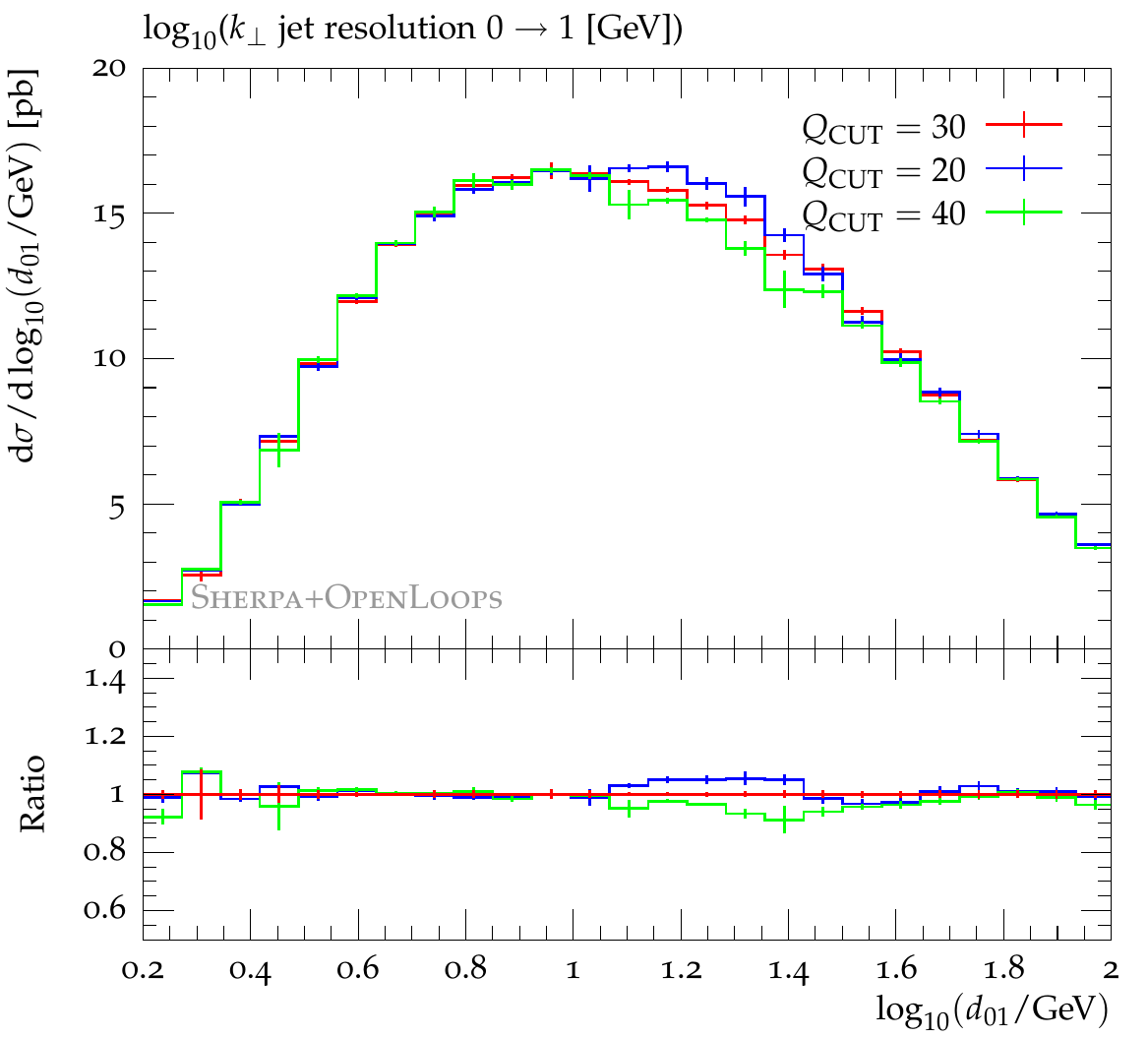}}
  \subfloat[$k_\perp$ jet resolution $1 \rightarrow 2$ \label{subfig:qcut_b}]{\includegraphics[width=0.33\linewidth]{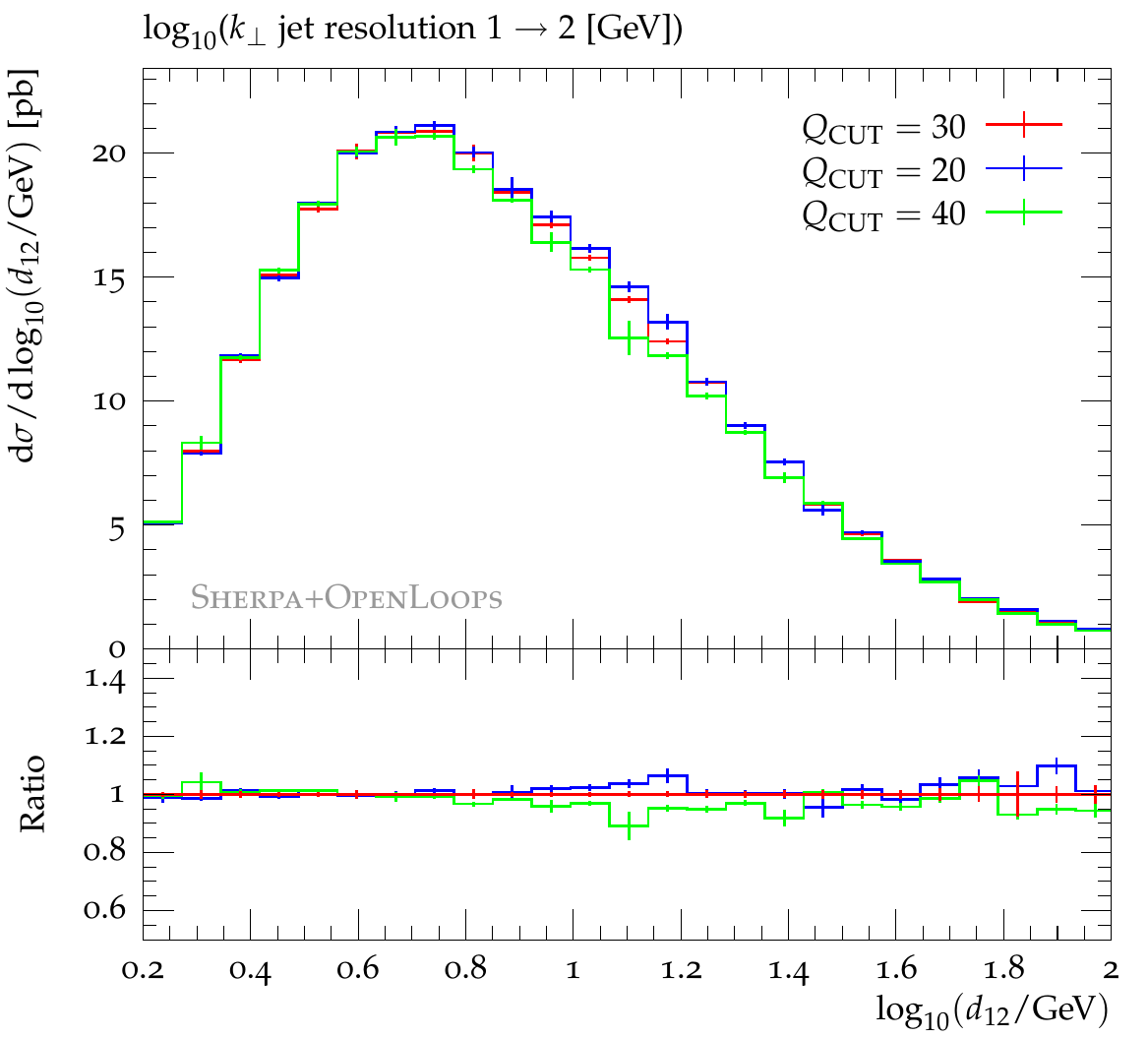}}
  \subfloat[$E_\perp^\gamma$ spectrum \label{subfig:qcut_c}]{\includegraphics[width=0.33\linewidth]{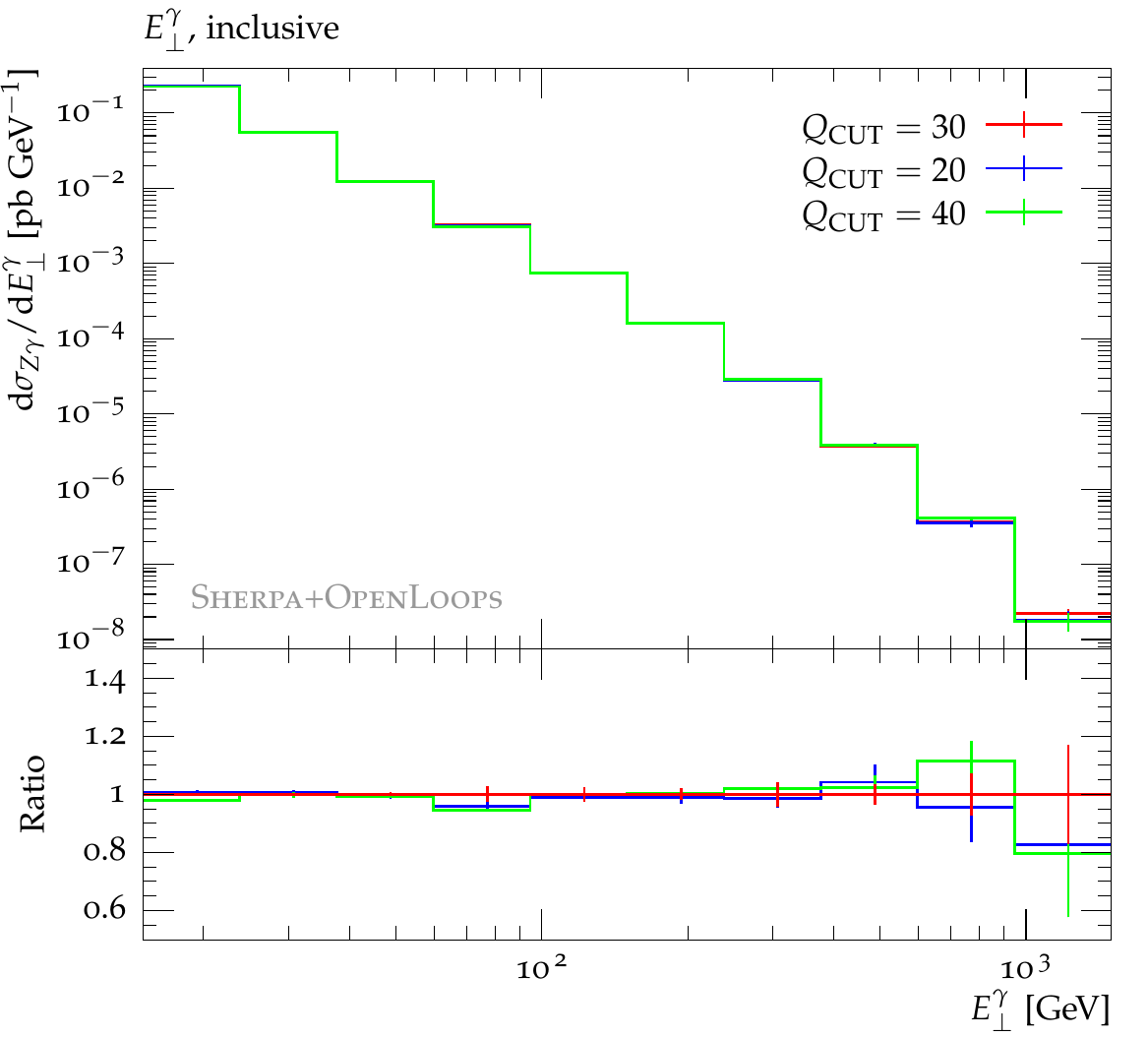}}
   \caption{Merging cut variation for $e^+ e^- \gamma + 0,1\mathrm{j@NLO} +
2,3\mathrm{j@LO}$ at 13\tev. The splitting scales are determined without any
additional phase space cuts, the generation cuts at matrix element level are
$p_\perp^\gamma> 15$\gev, mass$(e^+,e^- )>40$\gev, $\mathrm{\Delta} R (\gamma,
e^{\pm})>0.4$ and an smooth isolation cone with $R=0.1, n=2, \epsilon=0.1$. The phase space cuts used for the $E_\perp^\gamma$ spectrum are defined in Section~\ref{subsec:results_13_dir}. The error bars describe statistical fluctuations.}
   \label{fig:qcut}

\end{figure*}

Using the ME+PS merging method defined in Section~\ref{subsec:matching_merging} a new
parameter  is introduced, the merging cut $Q_{\mathrm{CUT}}$. It
separates the different phase space regions for the parton shower and higher
multiplicity matrix elements. Since this parameter is unphysical, physical
observables should be independent of its exact value as long it is chosen in a
reasonable range.

An interesting observable for checking this behaviour are the 
splitting scales as defined by the $k_\perp$-algorithm~\cite{Catani:1993hr}. All final state partons\footnote{In this section the simulation 
is performed at parton level for better scrutiny, i.e.\ multiple parton interactions and fragmentation are switched off.} 
are clustered to jets according to this algorithm. The splitting scale $d_\mathrm{(n-1)(n)}$ is than defined as the jet measure  which describes the cluster step of a $n$-parton final state to a $(n-1)$ final state.
Following this, $d_{01}$ gives the $p_\perp$ of the hardest jet and $d_{12}$ either describes the production of a second jet or the second splitting of the first jet in case there is only one. 

In the context of matching and merging, jets can emerge either from higher multiplicity matrix elements or from the parton shower.
Since the $k_\perp$ algorithm uses a jet measure which is very similar to
the jet criterion used by the merging algorithm, its splitting scales are very sensitive observables to study the interplay between LO/NLO matrix elements and partonic showers.

In Figure~\ref{fig:qcut}, these differential jet rates are evaluated for the first two splittings while varying the merging cut between 20 and 40\gev.
Figure~\ref{subfig:qcut_a} shows the hardest splitting scale. This jet rate is  sensitive to the transition from  the zero-jet NLO matrix element\footnote{A $n$-jet matrix element refers to a matrix element with $n$ additional, well separated partons in the matrix element.} with a shower emission at $Q_\mathrm{emission} < Q_\mathrm{CUT}$ 
to a one jet NLO matrix element with $Q_\mathrm{emission} > Q_\mathrm{CUT}$. 

By contrast, in Fig.~\ref{subfig:qcut_b} the subleading splitting scale is shown.
This splitting probes the transition from the one-jet NLO matrix element with an unresolved emission to a LO matrix element with two resolved jets. 

In both cases the uncertainties coming from the merging cut variation are less than 10\%. As mentioned above these splitting scales are shown since they are expected to be very sensitive to $Q_\mathrm{CUT}$. Indeed, other observables have a much smaller merging cut uncertainty. The $E_\perp^\gamma$ spectrum is given in Figure~\ref{subfig:qcut_c}, here the merging cut variation is found to be completely negligible in contrast to the scale uncertainties studied in Section~\ref{subsubsec:scalevar_13tev}. 

\FloatBarrier

\subsubsection{QCD core scale choice}
  
\begin{figure*}
  \centering
  \subfloat[$E_\perp^\gamma$ spectrum]{\includegraphics[width=0.5\linewidth]{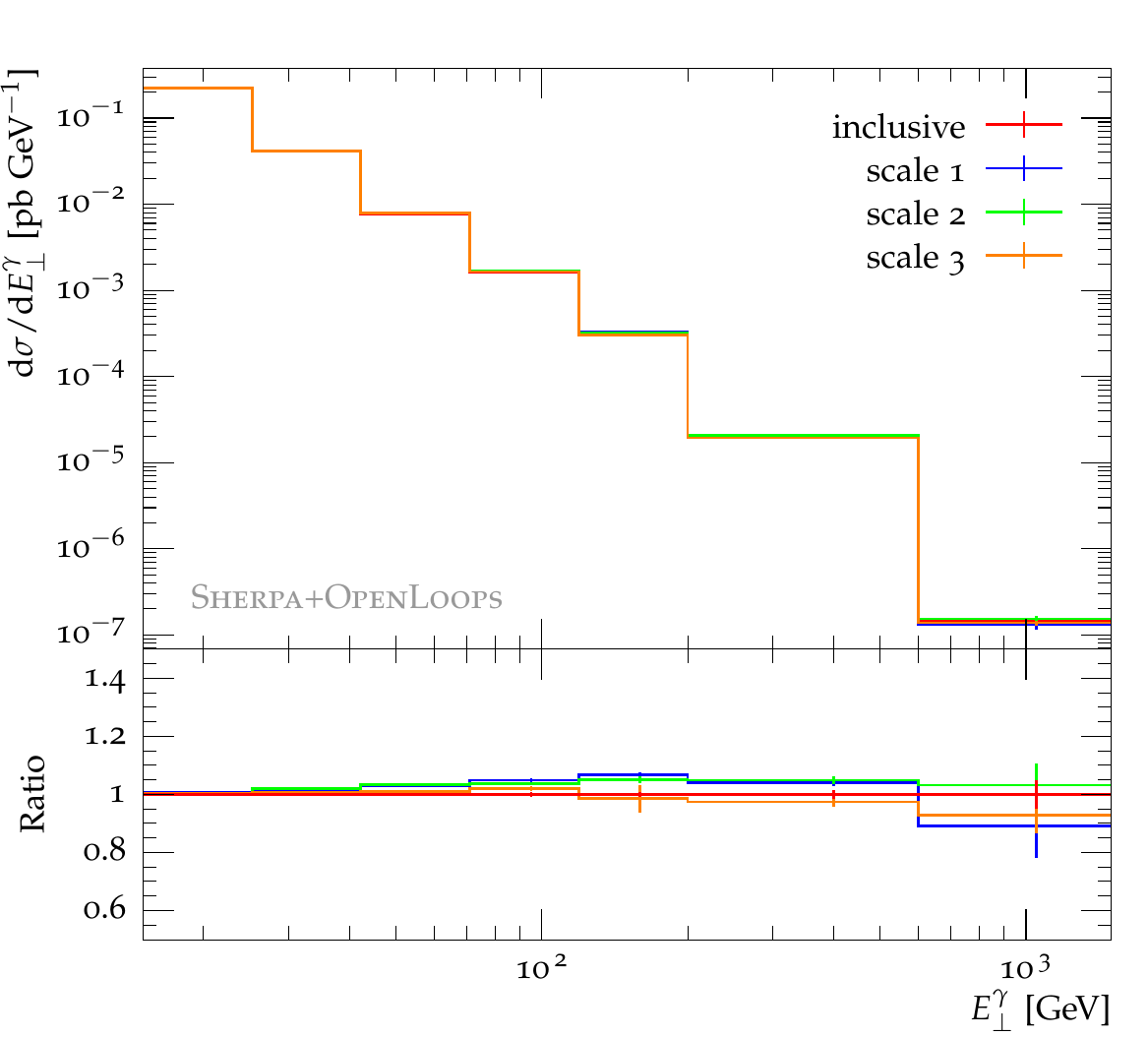}}
  \subfloat[$p_\perp$ distribution of the leading jet]{\includegraphics[width=0.5\linewidth]{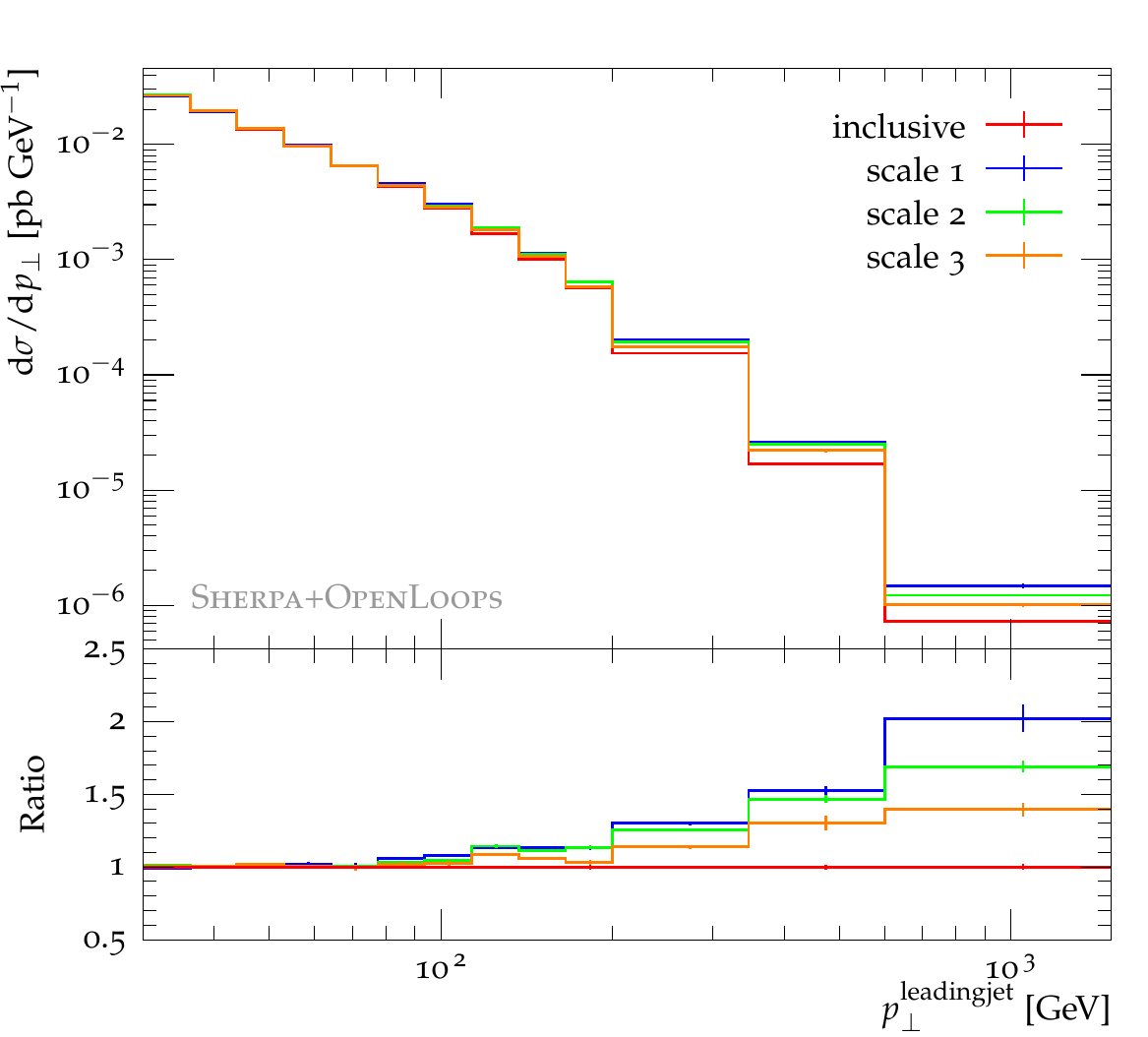}} 
  \caption{Comparison of different scale choices for a $pp\rightarrow e^+ e^- \gamma + 0,1j$@NLO + $2,3j$@LO setup.  \textit{inclusive} uses the inclusive clustering algorithm. 
  \textit{scale 1} and \textit{scale 2} correspond to the core scales defined in  Eq.~\eqref{eq:scale_1} and Eq.~\eqref{eq:scale_2} using exclusive clustering.
  \textit{scale 3} uses the exclusive clustering algorithm but the default core scale described in Eq.~\eqref{eq:def_core_scale}.}
  \label{fig:13tev_core_scales} 
\end{figure*}

After validating the stability of the MEPS@NLO method, here and in the next section the scale choices and variations will be discussed for a centre of mass energy of 13 \tev.
As before, all predictions are performed at shower level, meaning that hadronisation and multiple interactions are switched off explicitly. 
All phase space cuts for the 13 \tev\ analysis are summarized in Table~\ref{tab:cuts13tev}.

\begin{table}[]
\centering
\begin{tabular}{|c|c|} \hline
Lepton & $p_\perp>25$\gev, $\abs{\eta}<2.5$   \\
Jet    & $E_\perp>30$\gev, $\abs{\eta}<4.4$, $\DELTA 
R(\mathrm{jet}$, $e/\gamma ) > 0.3$       \\
Boson  & $M_{e^+, e^-}>40$\gev             \\
Photon & $E_\perp>15$\gev, $\abs{\eta}<2.5$ \\
Isolation & $\DELTA R(\gamma, e^\pm)>0.4$, $\epsilon_{\mathrm{iso}}=0.5$  \\
\hline
\end{tabular}
\caption{This table summarizes all cuts which define the
differential cross section analysis for 13\tev. The leading photon has to be isolated from all
other particles by fulfilling the requirement $\sum_{\DELTA R<0.4} E < \epsilon_{\mathrm{iso}}E^{\gamma}$ where the sum includes all particles which
have an angular distance of 0.4 or less to the photon axis. In addition, the photon is required to not come from a hadron decay. The leptons are 
dressed with all photons not coming from a hadron decay and within $\DELTA R < 0.1$.}
\label{tab:cuts13tev}
\end{table}

Different scale choices have been employed for $V\gamma$ production in the
literature. Two of them are 

\begin{equation}
\label{eq:scale_1}
\mu = \sqrt{{M_\mathrm{V}}^2 + {p_\perp^\gamma}^2 }
\end{equation}
 
and 

\begin{equation}
\label{eq:scale_2}
\mu = \sqrt{0.5 \left( {M_\mathrm{V}}^2 + {p_\perp^\gamma}^2 +  {(p^{e^+}+ p^{e^-})_\perp}^2 \right) }. 
\end{equation}

The former was used for the NNLO-QCD calculation of $V+\gamma$ in~\cite{Grazzini:2015nwa}, 
the latter is inspired by~\cite{Dixon:1999abc} and was used e.g.  in the NLO QCD+EW calculation~\cite{Denner201518} 
in a slightly modified manner.

Since this is a merged calculation it is not possible to use these scale definitions directly as described in Sec.~\ref{subsec:matching_merging}.
 However, the \texttt{STRICT\_METS} scale setter also allows to use custom scales for the core process.
In order to use this possibility, the clustering has to be restricted such that it results in a $Z\gamma$ core process. 
This can be enforced by using an \emph{exclusive} cluster mode which exactly reconstructs a possible shower history using QCD splittings only. 
If an allowed history is found, the core process is always $pp \rightarrow e^+ e^- \gamma$. 
It might also happen, that no possible shower history can be found, e.g. if all possible histories are unordered in terms of the shower variable. In such a case the remaining $pp \rightarrow e^+ e^- \gamma + n ~ \mathrm{partons}$ process is retained as core process.    
In both cases the remaining core process is finally evaluated using the core scales defined above. 

Altogether, four different scale choices are compared. The first one is "inclusive", it uses the default settings of \textsc{Sherpa} as described in Section~\ref{subsec:setup}.  
By contrast, the three remaining setups make use of the exclusive cluster mode introduced above and use different core scales as defined in Fig. \ref{fig:13tev_core_scales}.

In Figure~\ref{fig:13tev_core_scales} the differential $E_\perp^\gamma$ distribution and the $p_\perp$ spectrum of the leading jet are compared for the four different scale schemes.

In the $E_\perp^\gamma$ spectrum all scale choices are in good agreement and differ by not more than 10\%. Switching from the default to the exclusive clustering algorithm does not make a large difference for this observable. This also holds for all other observables which where measured by ATLAS and are discussed later on in Section~\ref{subsec:results8tev}, those measurements do not allow us to discriminate between the different scale choices. 

By contrast, the difference is much higher when looking at the $p_\perp$ spectrum of the leading jet. 
Here, all scale and cluster choices are in good agreement for low $p_\perp$ but differ as soon as $p_\perp$ exceeds 100\gev. At large values at order of 1000\gev\ the difference reaches almost a factor of two. 
There, the highest cross section corresponds to the core scale defined in Eq.~\eqref{eq:scale_1} in combination with the exclusive cluster model, whereas the lowest cross section is given by the default settings.   
This is not surprising since such a high $p_\perp$ region probes configurations where the jet is harder than the typical scale of this process, e.g. the $Z$ mass. 
Such configurations are unordered in terms of the parton shower evolution variable and are thus very sensitive to the clustering definition. 
If here a $pp \rightarrow Z \gamma + n ~\mathrm{partons}$ core process is determined but the core scale is evaluated solely based on $Z$ and $\gamma$, this scale will underestimate the physical scale and thus overestimate the strong couplings, resulting in a larger cross section.

However, based on the available information, there is no clear way to decide which cluster / scale settings are best suited for this process and the inclusive cluster settings are retained. Measuring the leading jet $p_\perp$ distribution in $Z\gamma$ events could greatly help to improve this situation.  

\subsubsection{Scale and PDF variation uncertainties}
\label{subsubsec:scalevar_13tev}

\begin{figure}
  \includegraphics[width=\linewidth]{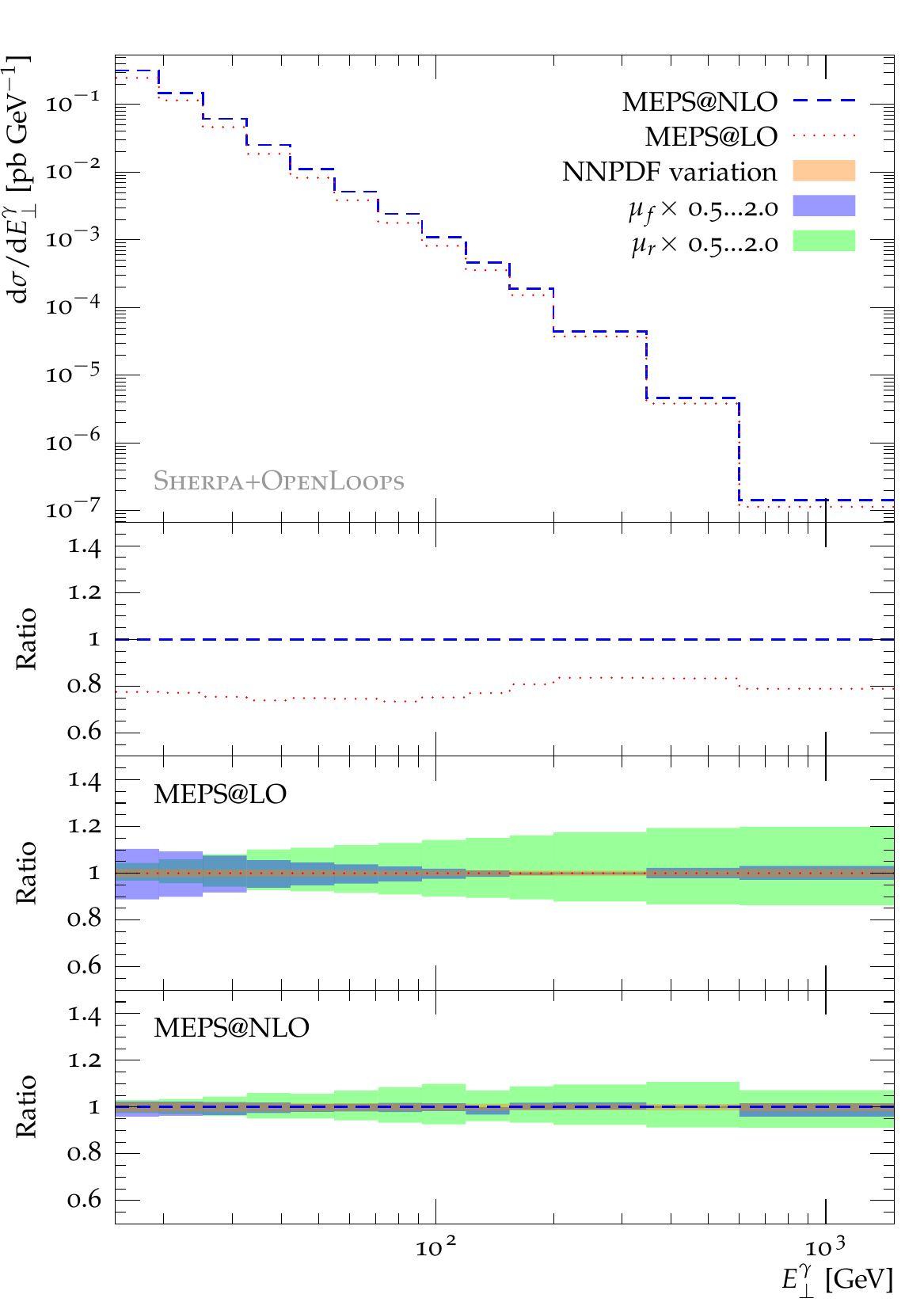}
  \caption{13\tev\ predictions for $E_\mathrm{T, inclusive}^\gamma$. }
  \label{fig:13tev_scale_et} 
\end{figure}

\begin{figure}
  \includegraphics[width=\linewidth]{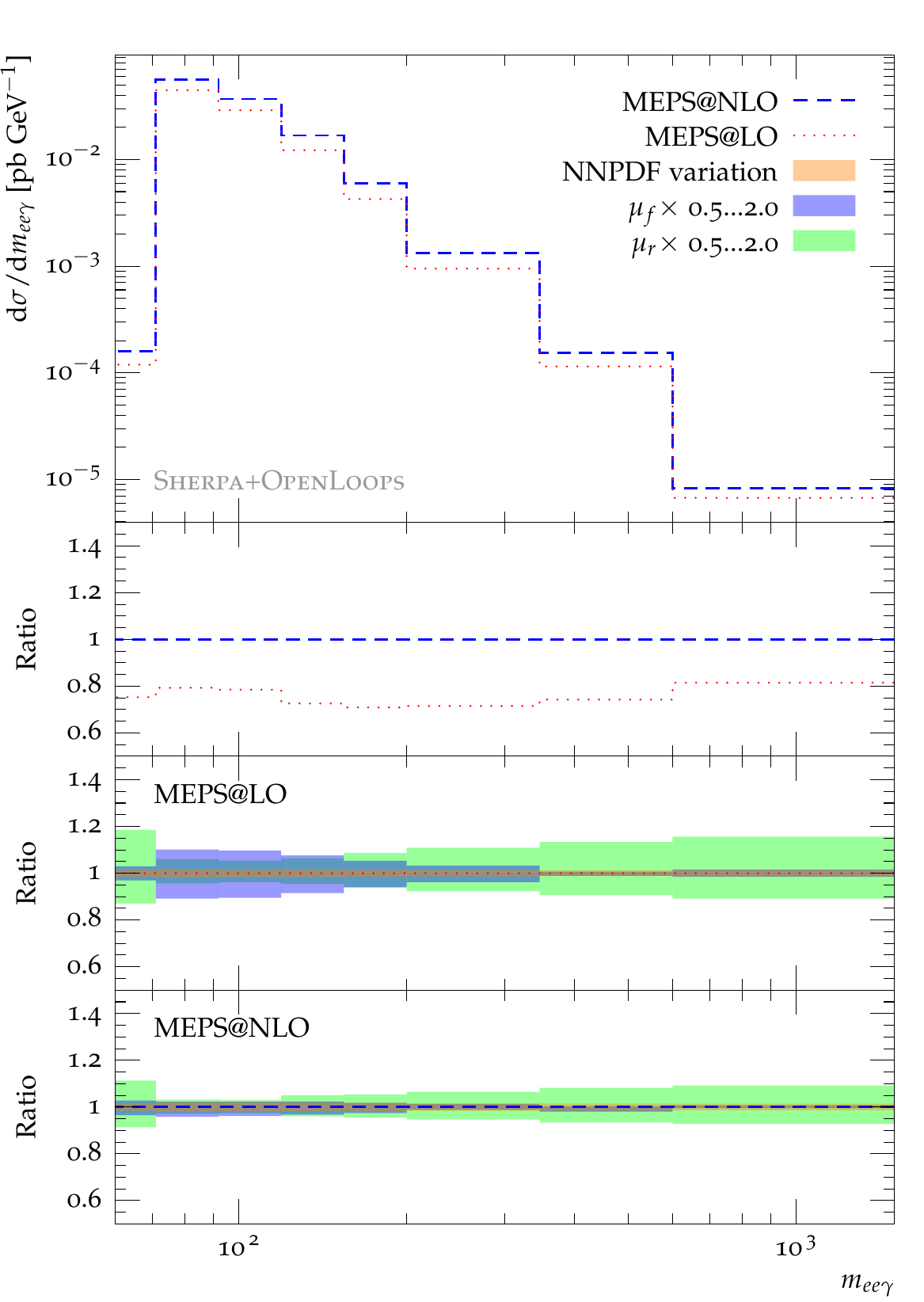}
  \caption{13\tev\ predictions for $m_{e^+ e^- \gamma}$. }
  \label{fig:13tev_scale_mzg} 
\end{figure}

\begin{figure}
  \includegraphics[width=\linewidth]{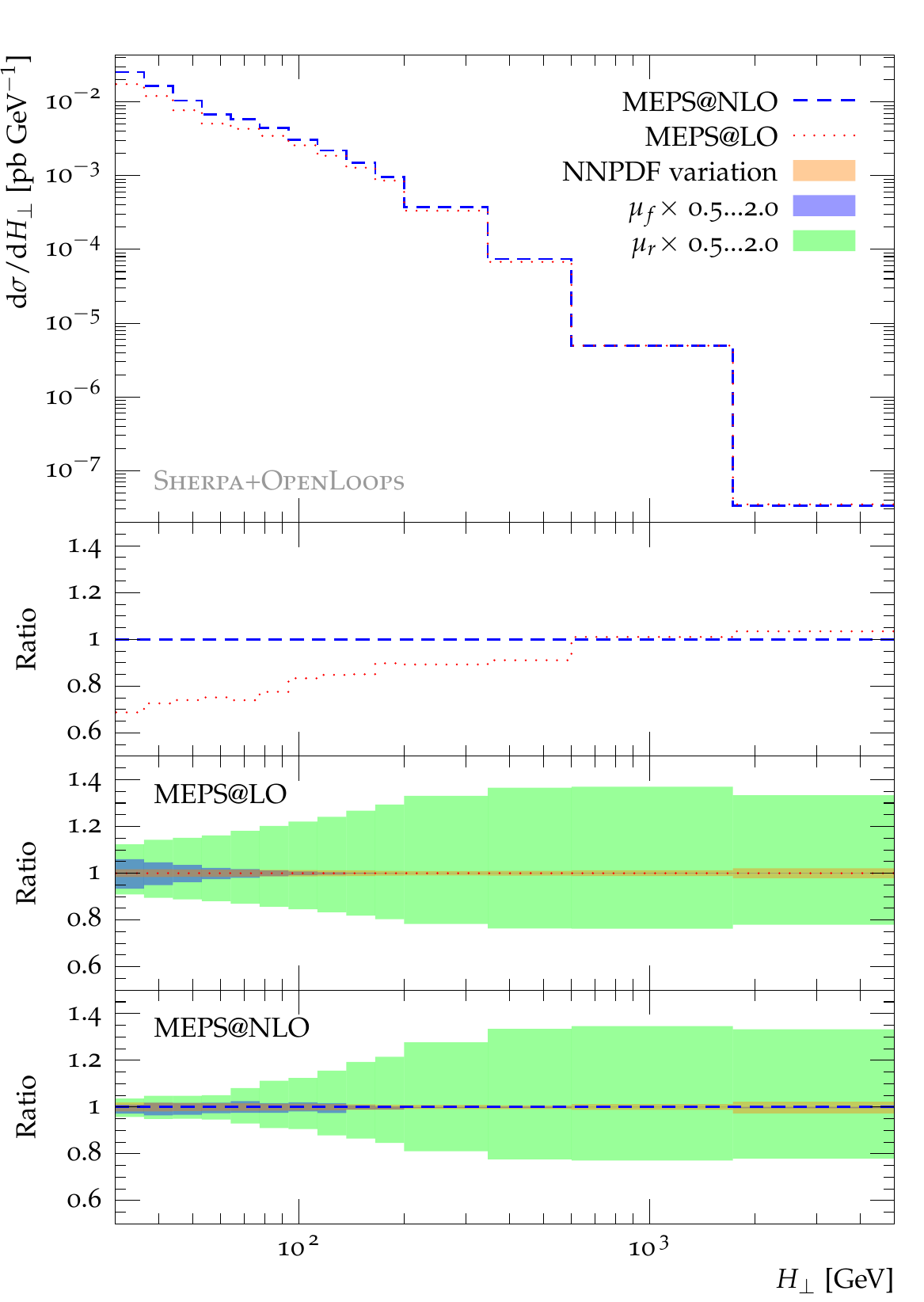}
  \caption{13\tev\ predictions for $H_\perp=\sum_{\mathrm{jets}}p_\perp^\mathrm{jet}$. }
  \label{fig:13tev_scale_ht} 
\end{figure} 

\begin{figure}
  \includegraphics[width=\linewidth]{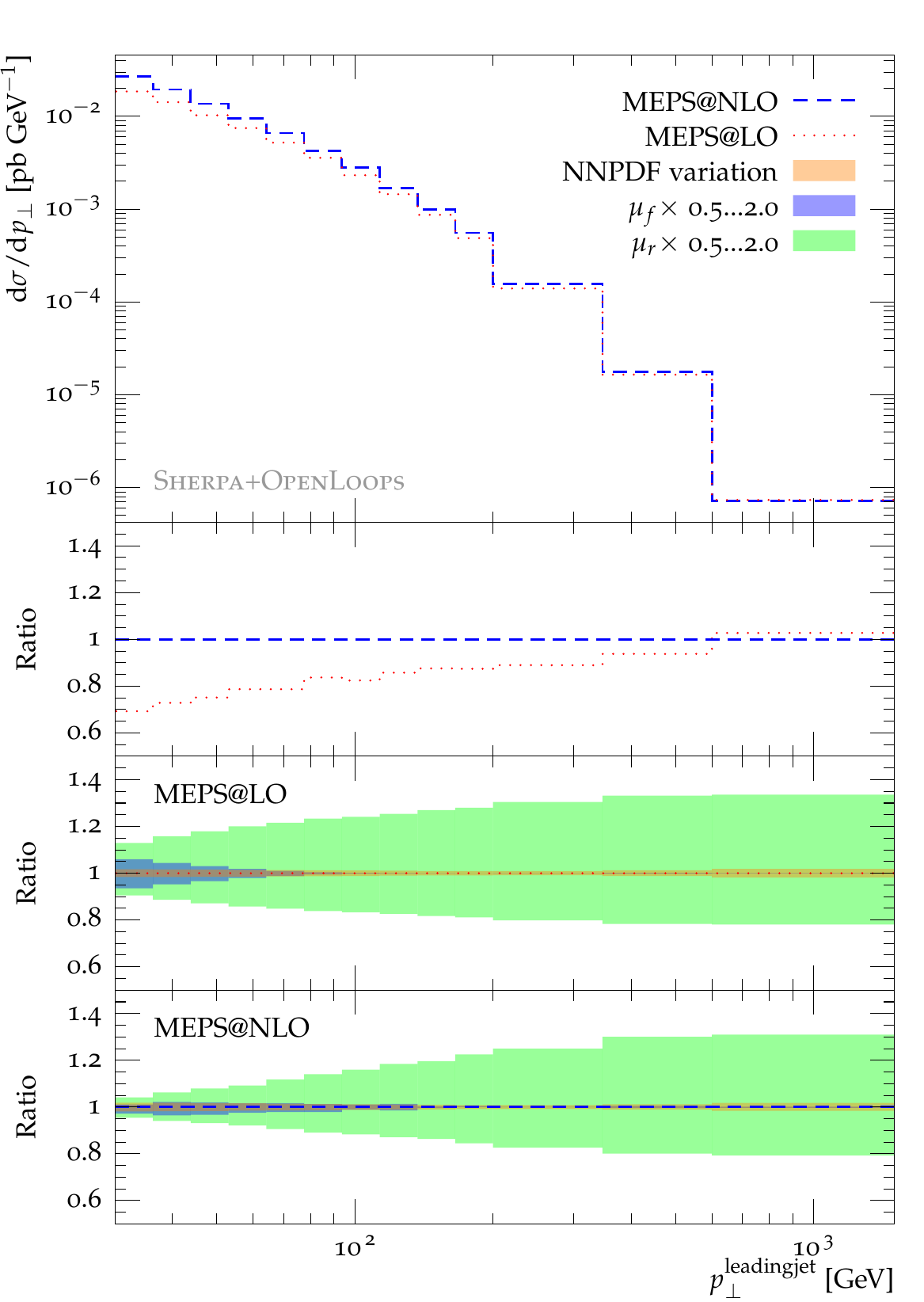}
  \caption{13\tev\ predictions for $p_\perp^\mathrm{leading jet}$.}
  \label{fig:13tev_scale_ptjet} 
\end{figure}

\begin{figure}
  \includegraphics[width=\linewidth]{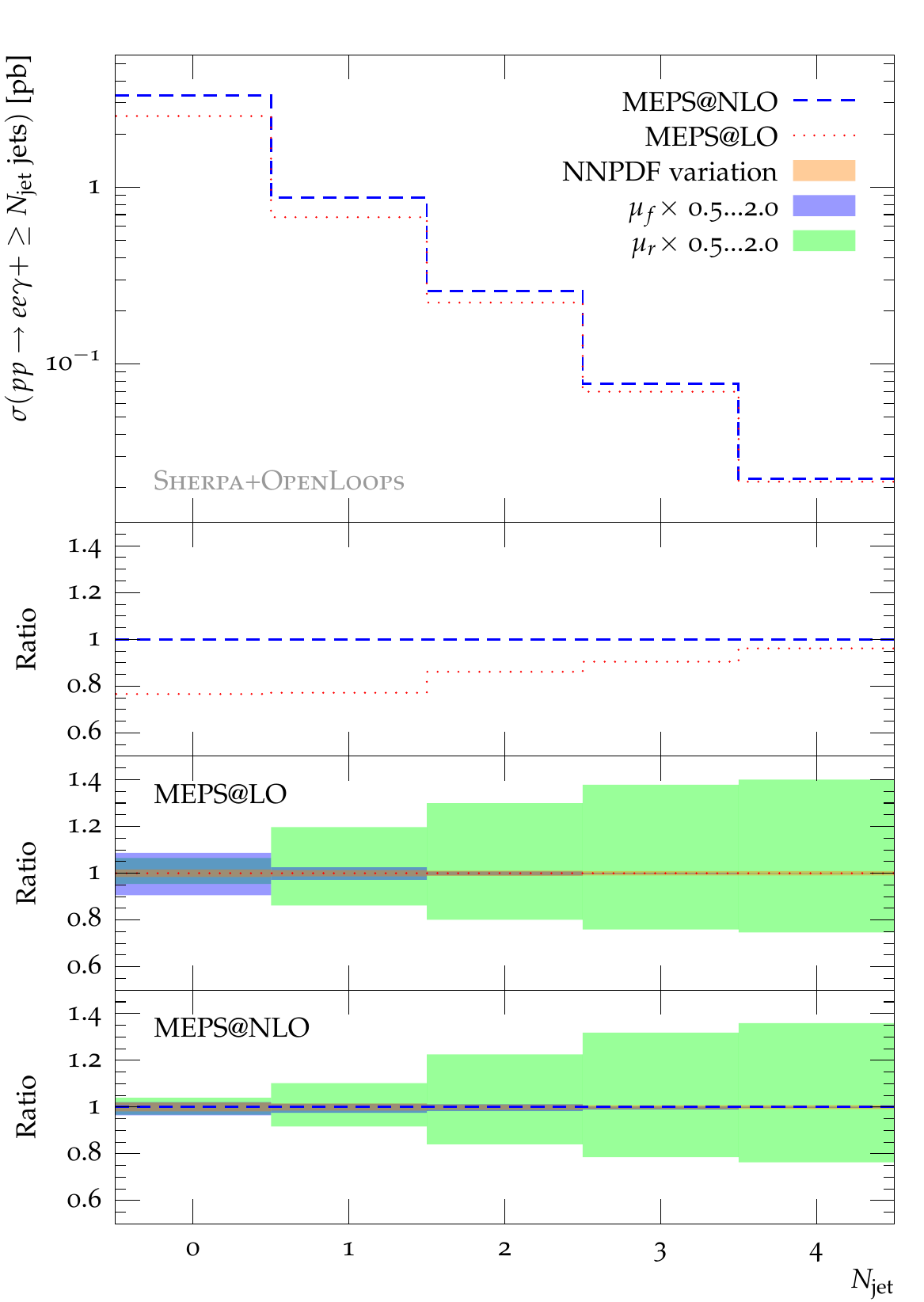}
  \caption{13\tev\ predictions for the jet multiplicity distribution. }
  \label{fig:13tev_scale_njet} 
\end{figure}

In addition to the core scale variations studied above, independent variations of the $\mu_\mathrm{R}$ and $\mu_\mathrm{F}$ scales are performed. All replicas of the NNPDF set are used to estimate the PDF uncertainty. These variations include only contributions from the matrix elements but not from the shower. 

All figures are structured as follows.
Each figure has three ratio plots. 
The main plot and the first ratio plot point out the difference between MEPS@NLO and MEPS@LO. Here, the MEPS@NLO prediction is choosen as reference.
By contrast, the additional subplots show the size of all performed scale variations for each method as a ratio with respect to the corresponding nominal prediction.

Figures~\ref{fig:13tev_scale_et} and \ref{fig:13tev_scale_mzg} show predictions for $E_\perp^\gamma$ and $M_{ll\gamma}$. 
The corrections between MEPS@LO and MEPS@NLO are almost flat at a level of $20\%$.
As expected, the scale uncertainties are reduced when moving from MEPS@LO to MEPS@NLO. 
The factorisation scale dependency vanishes almost completely for $p_\perp^\gamma \lesssim 60\gev$ and $m_{ll\gamma}$ close to the $Z$ peak, compared to 10\% in the MEPS@LO case. 
The renormalisation scale uncertainty is reduced by roughly a factor of two for both observables, it shrinks from 20\% to 10\% for high $E_\perp^\gamma$.

Figure~\ref{fig:13tev_scale_ht} shows $H_\perp$, which is defined as the sum of the transverse momenta of all jets fulfilling the conditions defined in Table~\ref{tab:cuts13tev}, $H_\perp = \sum_\mathrm{jets} p_\perp^\mathrm{jet}$.

Here both the factorisation and renormalisation scale uncertainties are reduced significantly for small $H_\perp$ but have almost the same size if $H_\perp$ exceeds $200\gev$. 
The leading jet $p_\perp$ is depicted in Fig.~\ref{fig:13tev_scale_ptjet} and shows a very similar behaviour. 
Again the renormalisation scale uncertainty is reduced from 15\% to 5\% for low $p_\perp^\mathrm{jet}$ but does not change for values from 200\gev\ onwards. 
Finally, Fig.~\ref{fig:13tev_scale_njet} shows the jet multiplicity. 
In the zero-jet bin the factorisation scale dominates when using MEPS@LO, this uncertainty vanishes almost completely when moving to MEPS@NLO. However, the one jet bin is dominated by the renormalisation scale unertainty. This uncertainty is reduced from 20\% to 10\% when moving to MEPS@NLO.

All of these observables only show minor improvements in the multi-jet regions.
This is not surprising since observables which are sensitive to a high number of hard jets are hardly improved by the MEPS@NLO applied here. Only the zero- and one-jet matrix elements are calculated at next-to-leading order but the two- and three-jet calculations still have leading order accuracy. Both $H_\perp$ and $p_\perp^\mathrm{jet}$ are dominated at large values by multi-jet configurations and are thus described at leading order accuracy only.  
This effect is reflected by the ratio between the MEPS@NLO and the MEPS@LO method, too. At low values of $H_\perp$ and $p_\perp^\mathrm{jet}$ it amounts to 0.7 but increases to one for larger values.

\FloatBarrier


\subsection{Comparison with $\sqrt s=$8\tev\ measurements}
\label{subsec:results8tev}

\begin{figure}
 \includegraphics[width=\linewidth]{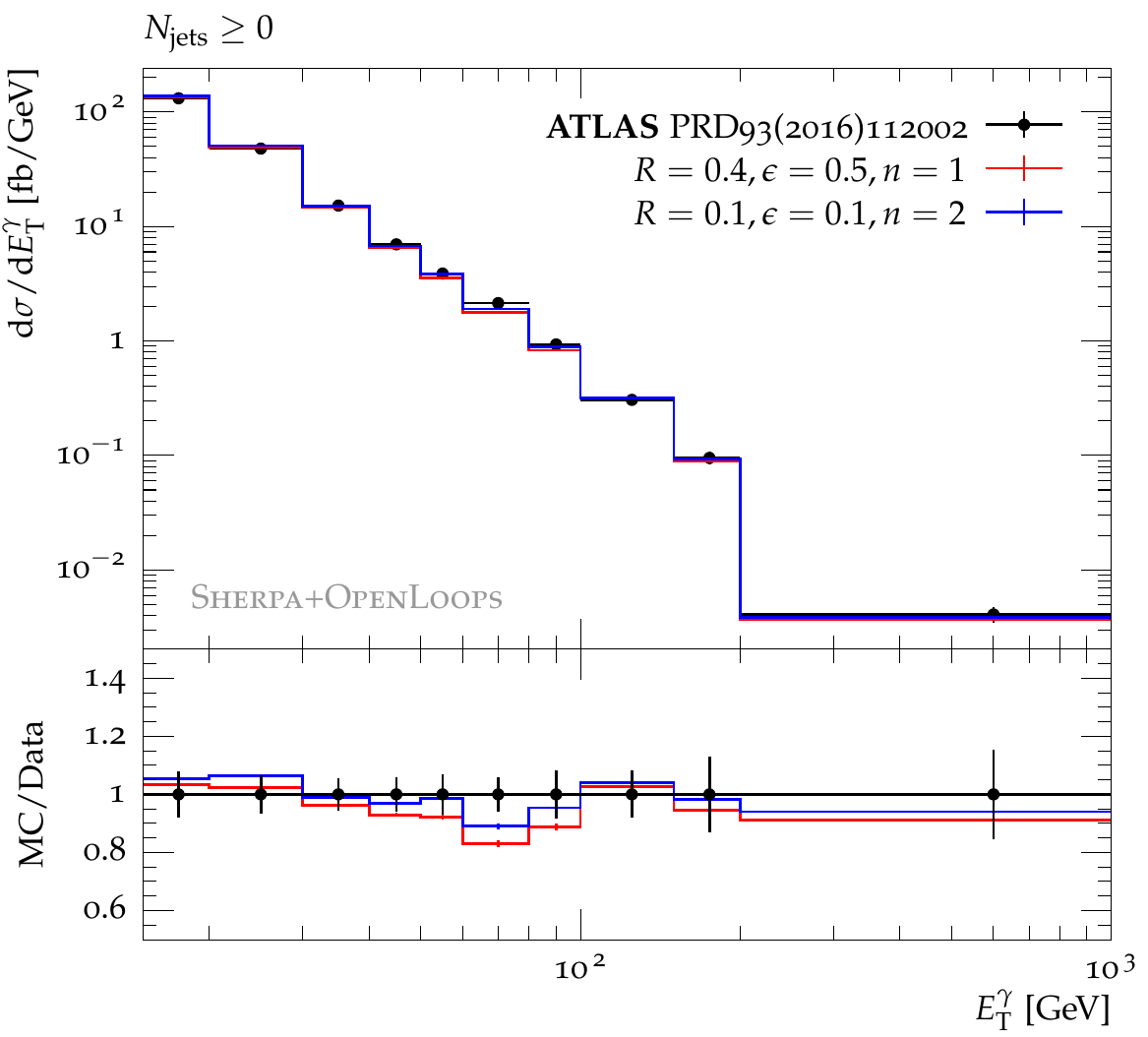}
 \caption{Comparison of two different parameter sets using
the smooth isolation criterion at a centre of mass energy of 8\tev. 
}
 \label{fig:iso_var_8}
\end{figure}

\begin{figure*}
  \centering
  \subfloat[$pp \rightarrow e^+ e^- \gamma + \mathrm{jets}, ~ N_\mathrm{jets} \ge 0$]{\includegraphics[width=0.5\linewidth]{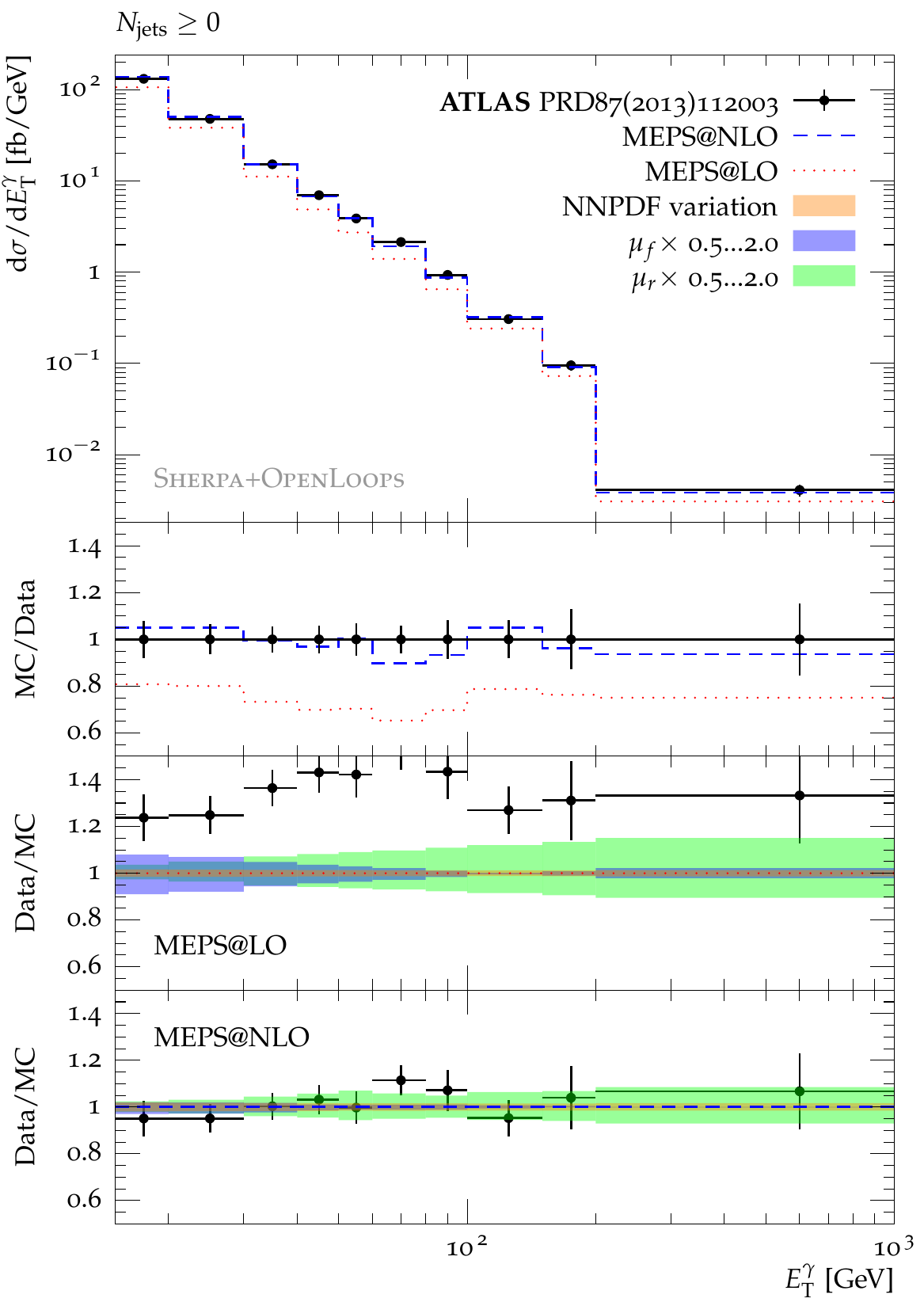}}
  \subfloat[$pp\rightarrow e^+ e^- \gamma, ~ N_\mathrm{jets} = 0 $]{\includegraphics[width=0.5\linewidth]{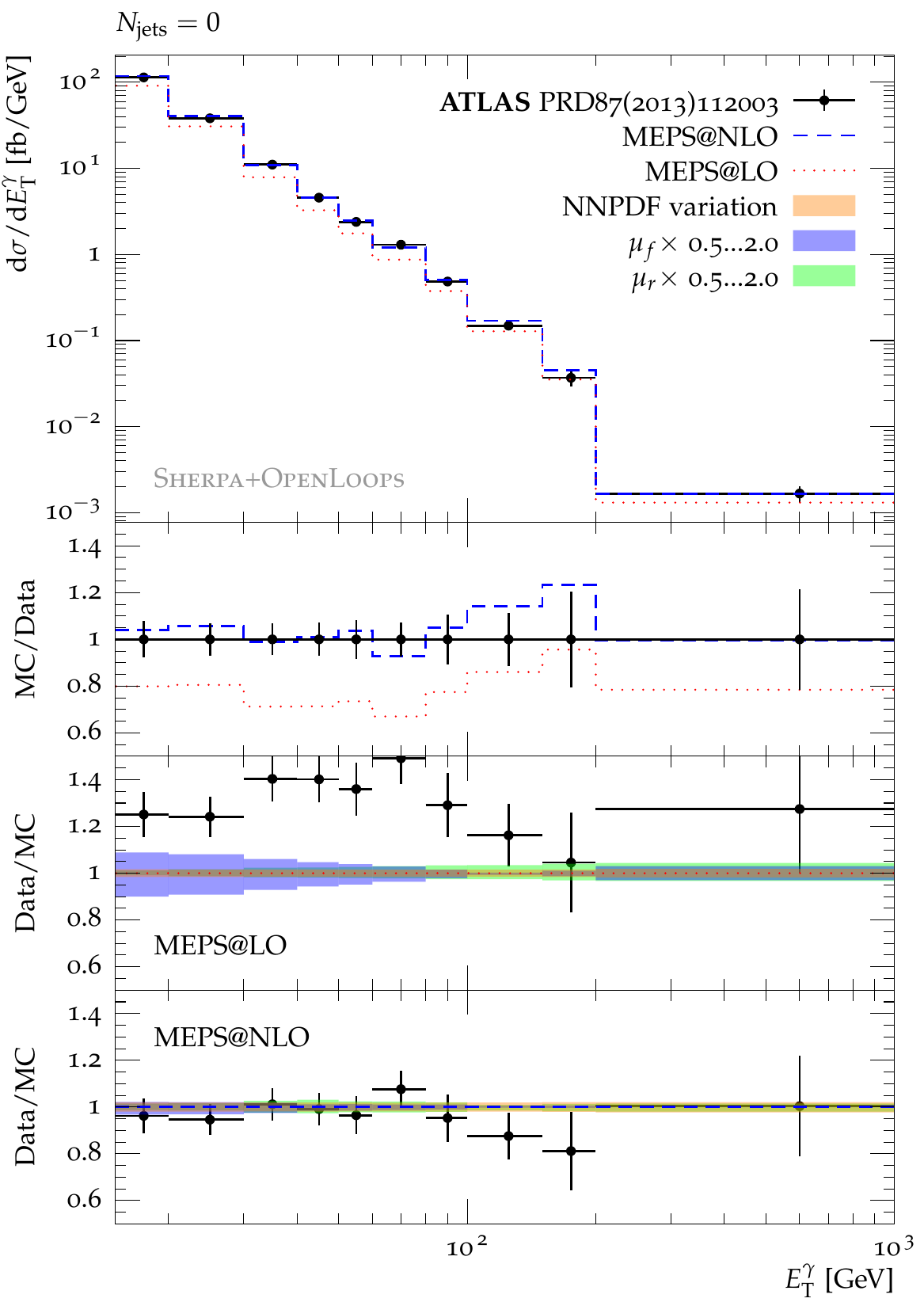}}
  \caption{$E_\perp^\gamma$ spectrum measured by ATLAS in $pp\rightarrow e^+e^- \gamma$ at 8\tev, compared for MEPS@LO and MEPS@NLO.}
   \label{fig:8tev_scale_var_et}
\end{figure*}

\begin{figure*}
  \centering
  \subfloat[$pp \rightarrow e^+ e^- \gamma + \mathrm{jets}, ~ N_\mathrm{jets} \ge 0$]{\includegraphics[width=0.5\linewidth]{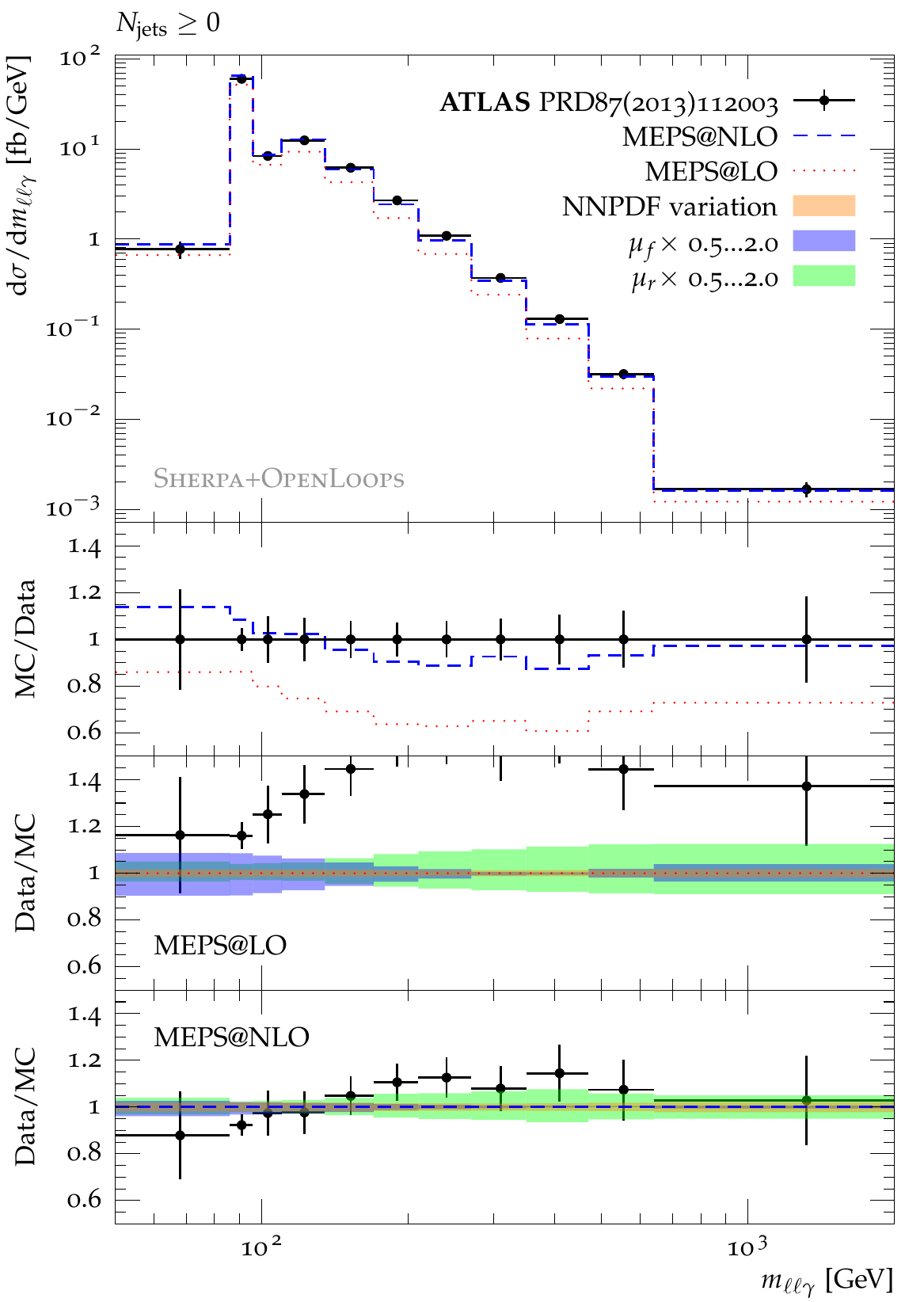}}
  \subfloat[$pp\rightarrow e^+ e^- \gamma, ~ N_\mathrm{jets} = 0 $]{\includegraphics[width=0.5\linewidth]{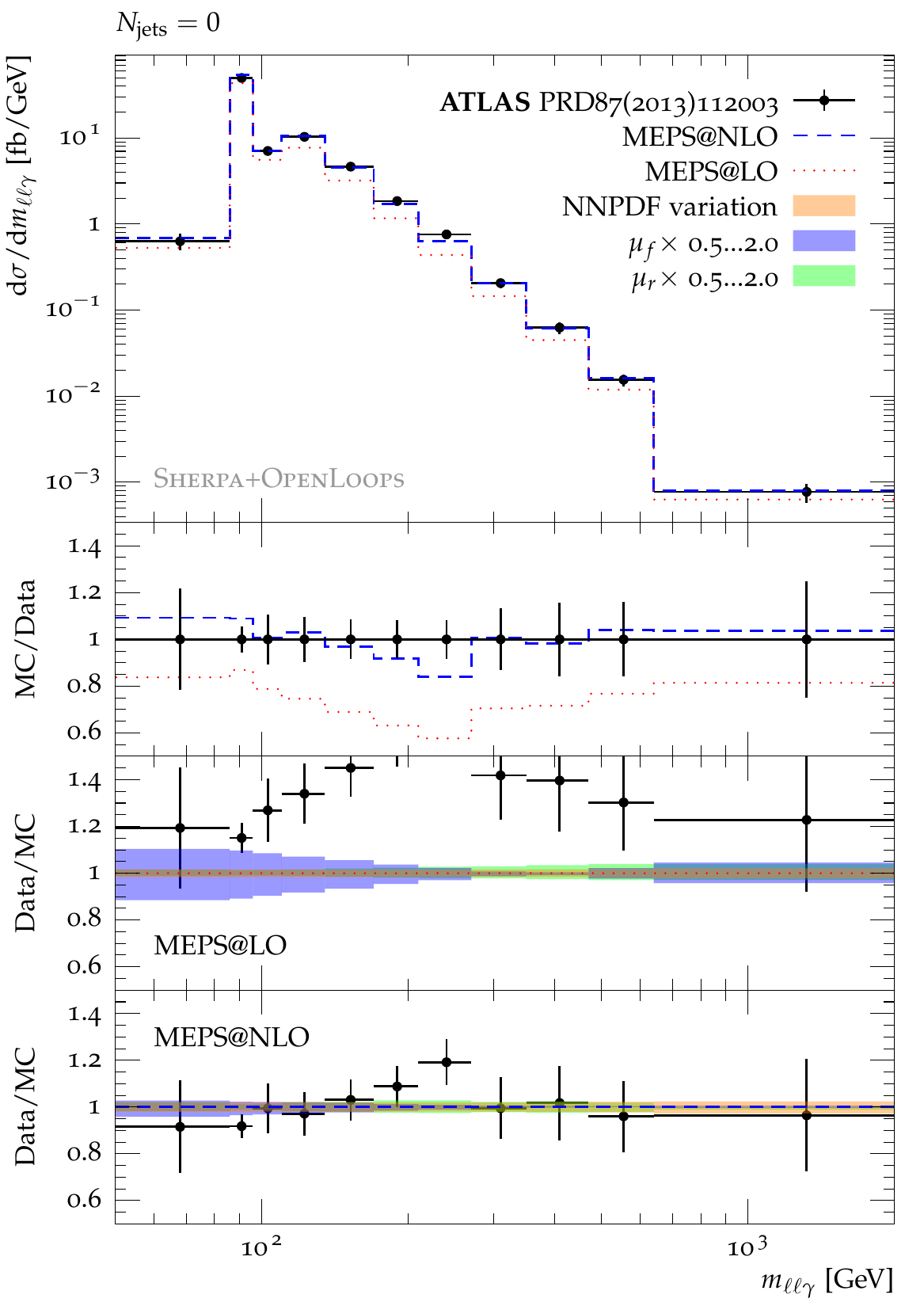}}
  \caption{$m_{ll\gamma}$ measured by ATLAS in $pp\rightarrow e^+e^- \gamma$ at 8\tev, compared for MEPS@LO and MEPS@NLO. }
   \label{fig:8tev_scale_var_m}
\end{figure*}

\begin{figure}
  \includegraphics[width=\linewidth]{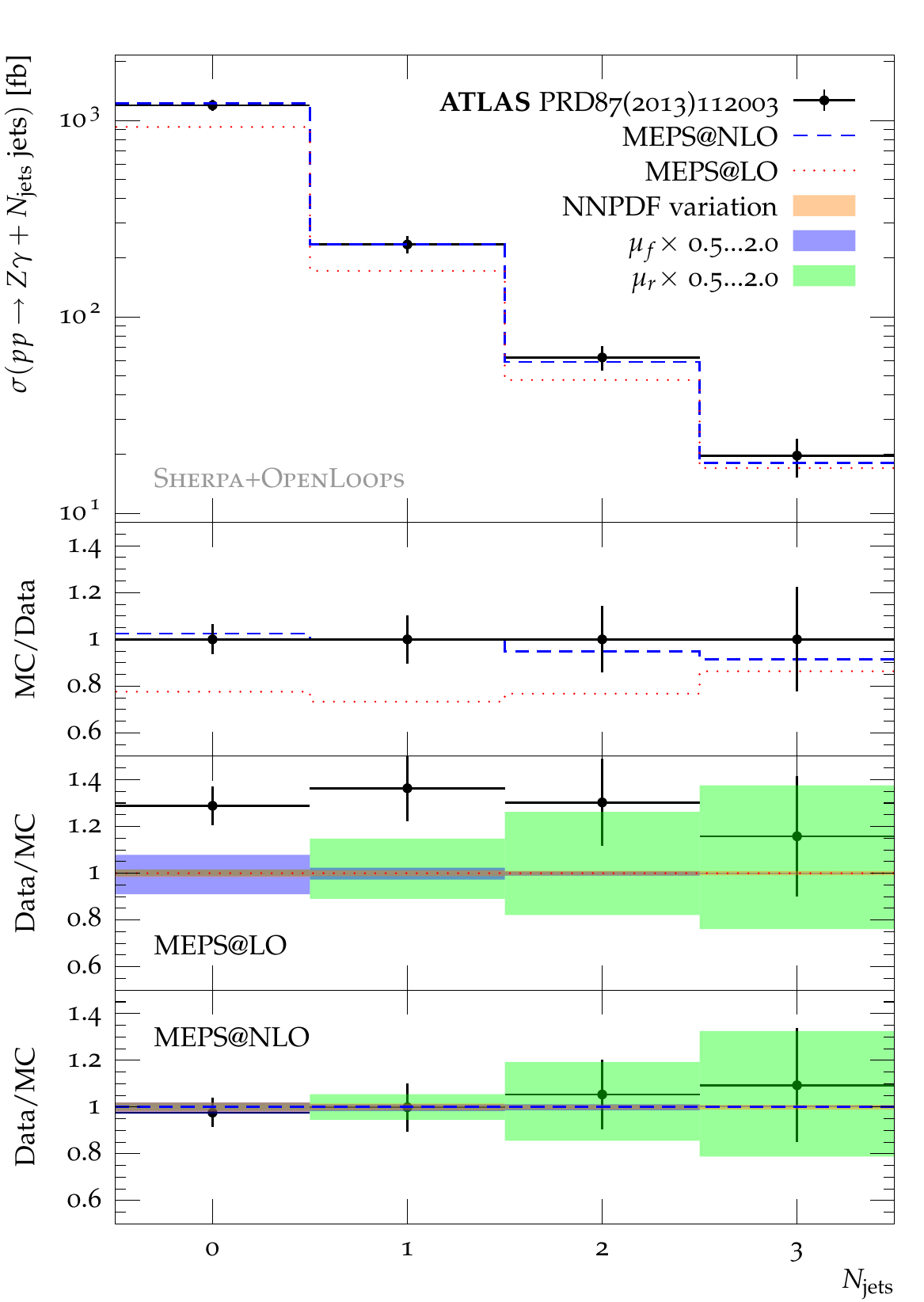}
  \caption{Jet multiplicity measured by ATLAS in $pp\rightarrow e^+e^- \gamma$ at 8\tev, compared for MEPS@LO and MEPS@NLO. }
   \label{fig:8tev_scale_var_njet}
\end{figure}

Both, the stability and the reduction of the perturbative uncertainities have been demonstrated in the 13 \tev\ results in the last section. Now, the focus is on the comparison with recent experimental data and a study of the interplay between the smooth isolation criterion with the experimental one.

This section relies on a measurement of the ATLAS collaboration at 8\tev~\cite{PhysRevD.93.112002}, using 20.3 fb$^{-1}$ of data.  In this measurement,
final states with $ll\gamma$  and up to three jets are studied. 
Here, the focus lies on the $e^+ e^- \gamma + \mathrm{jets}$ final state. All
cuts which define the extended, differential
fiducial cross section are summarized in Table~\ref{tab:atlascuts8tev}. 

The first topic to be studied is the isolation criterion used by this analysis. 
As described in Section~\ref{subsec:iso_photo} our prediction uses the smooth cone isolation criterion.
By contrast, the experimental isolation is based on anti-$k_\perp$-jets with $\DELTA R=0.4$. These jets include all particles except neutrinos and muons and are not required to fulfil the cuts described in Table~\ref{tab:atlascuts8tev}.
A photon is defined as isolated if either the nearest jet has an angular distance $\DELTA R>0.4$ to the photon axis or this jet's transverse energy $E_\perp^\mathrm{jet}$ fulfils
\begin{equation}
\frac{E_\perp^\mathrm{jet} - E_\perp^\gamma}{E_\perp^\gamma } = \epsilon_\mathrm{exp} < \epsilon_\mathrm{max}.
\end{equation}
Even if one assumes that both the final jet $E_\perp$ and its direction are in perfect agreement with the closest parton $E_\perp$ at matrix element level, this criterion differs from the one used in our calculation and described in Section~\ref{subsec:iso_photo} when going to lower angular distances. 
 
As a consequence, two different parameter sets for the smooth isolation criterion
are compared. The first set is based on the 2013 Les Houches report~\cite{Butterworth:2014efa}
which recommends the usage of the smooth isolation criterion for fixed order
calculations if the parameters are matched to the experiment. Following
this, the parameters are $R=0.4$, $n=1$ and $\epsilon=0.5$. However, since this is not a fixed
order calculation, the smooth isolation criterion is used only at matrix element level and in
the subsequent final state analysis the experimental cut has to be
passed additionally. 

The second parameter set is thus chosen more inclusively in $R$, here $R=0.1$, $n=2$ and $\epsilon=0.1$.
Such a setup also reflects the requirement that experiments want to generate event samples to be as
universal as possible, usable not only as signal process but also as background
for many other measurements. 

In Figure \ref{fig:iso_var_8} both these parameter sets are compared with the $E_\perp^\gamma$ spectrum measured by ATLAS.
Both predictions are in good agreement with the data but the more inclusive set gives a slightly higher cross section. 
This is most obvious in a $p_\perp^\gamma$ region of around 70\gev, there the difference reaches almost 10\%. 
Although it is expected that the first set with $R=0.4$ may miss some contributions due to the smoothing of the cone, it is not guaranteed that the second set gives a more accurate prediction. 
A more inclusive parton-level isolation  always  allows configurations which come closer to the collinear, non-perturbative region. This region cannot be described without fragmentation functions or QED parton shower matching~\cite{Hoeche:2009xc}.  However, this is not expected to happen if the angular distance is large enough  and thus the inclusive parameter set is used in the following.  

All measured observables are shown in Figure~\ref{fig:8tev_scale_var_et}-\ref{fig:8tev_scale_var_m}.
The plots are structured in the same way as in Section~\ref{subsubsec:scalevar_13tev}. 
Both methods, MEPS@LO and MEPS@NLO, are compared to the data and directly to each other.
In the main plot and the first ratio plot both the MEPS@LO and the MEPS@NLO predictions are compared to the measured data. In addition, two further ratio plots show the impact of the described perturbative variations. In contrast to the first ratio plot the respective nominal prediction is chosen as reference here since this simplifies a direct comparison of both methods.
 
In almost all observables the MEPS@NLO prediction is in excellent agreement with the data. A small deviation is found in the invariant mass prediction with zero jets in a region around 250 GeV.
The MEPS@NLO differential cross sections are about 20\% larger with respect to the MEPS@LO results at small scales. At large scales the difference gets smaller since the contribution of the additional LO jets is increasing.   

As already seen in the 13 \tev\ section, the uncertainties estimated by the scale variations are reduced when moving from MEPS@LO to MEPS@NLO. At MEPS@LO, for lower values of $E_\perp^\gamma$ and $M_{ee\gamma}$ the factorisation scale is the dominant source of uncertainty and reaches up to 10\% . This is reflected in the zero jet bin of Figure~\ref{fig:8tev_scale_var_njet}, too. By contrast, in the MEPS@NLO case this uncertainty is removed almost completely.

At higher values of $E_\perp^\gamma$ or $m^{ee\gamma}$ the renormalisation scale uncertainty takes over in all inclusive observables and reaches values of 10-20 \% in the MEPS@LO case. This uncertainty is reduced for MEPS@NLO  to 5-10\% in both large $E_\perp^\gamma$ and the lower jet multiplicity bins.

Both, the size of the corrections and their uncertainties behave very similarly between 8 and 13 \tev.

\begin{table}[]
\centering
\begin{tabular}{|c|c|} \hline
Lepton & $p_\perp>25$\gev, $\abs{\eta}<2.47$   \\
Jet    & $E_\perp>30$\gev, $\abs{\eta}<4.45$, $\mathrm{\Delta} R(\mathrm{jet}$,
$e/\gamma ) > 0.3$                   \\
Boson  & $M_{e^+, e^-}>40$\gev                                                  
                             \\
Photon & $E_\perp>15$\gev, $\abs{\eta}<2.37$ \\
Isolation & $\mathrm{\Delta}R(\gamma, e^\pm)>0.7$, $\epsilon_{\mathrm{max}}=0.5$  \\
\hline
\end{tabular}
\caption{This table summarizes all cuts which define the extended, fiducial
cross section in the measurement~\cite{PhysRevD.93.112002}. Leptons are dressed with all photons having an angular distance of $\mathrm{\Delta}R<0.1$.}
\label{tab:atlascuts8tev}
\end{table}

\FloatBarrier

\section{Interplay with $Z$+jets production and QED final state radiation}
\label{sec:results_fsr}

\subsection{Motivation}
\label{subsec:or_motivation}

When predictions for $V\gamma$ production are used in experimental searches
to determine background contributions, they have to be combined with
predictions for $V$+jets production in several cases. A jet from the $V$+jets
sample can be misidentified as a photon at the detector level and thus
contribute to the $V\gamma$ event selection. Another example is a selection
requiring multiple leptons, if the photon is misidentified as an electron.

At the same time the two types of MC samples are not exactly complementary:
the simulation of QED final state radiation (FSR) from the leptons in the
$V$+jets sample generates a fragmentation contribution also contained in the FSR-like
diagrams of the $V\gamma$ process. It is obvious that this overlap has to
be removed before the samples can be used for background estimation.

The overlap removal is a conceptually straightforward requirement, which is
complicated by two facts. The QED FSR photons in the $V$+jets simulation are
produced at the hadron level and can thus not simply be subjected to
parton-level cuts matching the ones in the $V\gamma$ simulation.
Furthermore the photon cuts in a multi-jet merged sample of $V\gamma$+jets
require an isolation of the photon with respect to partons from the multi-jet
matrix elements. This constraint has to be respected when defining the
complementary cuts for the $V$+jets sample.

An implementation of such an overlap removal at the event generation level is
discussed in this section using the example of $V=Z$.

\subsection{Implementation of overlap removal}
\label{subec:or_imp}
In order to combine $Z$ and $Z\gamma$ events\footnote{Here and in the following $ll(\gamma) + \mathrm{jets}$ final states are denoted as $Z(\gamma)$ for better readability.} the phase space is split into two regions. 
The $Z\gamma$ process includes photons directly in the matrix elements. The phase space of this region should be as large as possible but is limited since the matrix elements diverge when the photon is soft or collinear either to a massless lepton or quark\footnote{In this publication all leptons and quarks except the top are treated as massless in the matrix elements.}. By contrast, photons generated by YFS in $Z$ events do not have these limitations, but can not describe initial state radiation which usually gives most of the contribution to hard photons.

In principle, the phase space slicing is defined by three components. First, a $p_\perp^{\gamma}$ cut, secondly a lepton photon isolation and finally a photon hadron isolation. These cuts exclude a region where collinear or soft divergences are present and no fixed order calculation in QCD is possible. Thus, an $Z\gamma$ event has to pass all these cuts while an $Z$ event has to fail at least one of them.

In case of  $Z\gamma$ events, there are already cuts at matrix element level present and it would be desirable to use them directly for the overlap removal. Unfortunately, this is not possible since the generation of additional final state photons via YFS happens technically after the parton shower. The shower can shift the kinematics of all particles, thus cutting once before and once after the shower would result in a mismatch. As a consequence, the slicing cuts are applied to both $Z$ and $Z\gamma$ events at hadron level.

Hadron level cuts are not supported by \textsc{Sherpa} out-of-the-box. For this study, a support for custom modules was implemented which  makes it possible to veto events at the hadron-level very flexibly. This feature will be available within the next \textsc{Sherpa} release.

Special care has to be taken when selecting the photon and leptons which take part in the slicing procedure. Non-prompt leptons and photons can easily be produced by the decay of hadrons and there is no requirement that these further particles are softer than the particles coming directly from the hard interaction or the YFS algorithm. However, it has to be guaranteed that all divergences which are present at matrix element level are covered by the slicing cuts since otherwise the result would still depend on the matrix element level cuts. 

As a consequence, the hardest photon which does not come from a hadron decay is used for the definition of the slicing variables. The photon-lepton isolation is applied only to prompt leptons. For the hadronic isolation all particles excluding the prompt leptons and the photon are taken into account.

In the following, isolated photons are required to fulfil $p_\perp>10\gev$, a photon-lepton isolation of $\DELTA R > 0.4$ and the hadronic isolation using a smooth cone isolation with $R=0.4$, $n=1$ and $\epsilon=0.5$.

\subsection{Results}
\label{subec:or_results}

\begin{figure*}[p]
  \centering 
  \subfloat[inclusive $E_\perp$ spectrum]{\includegraphics[width=0.5\linewidth]{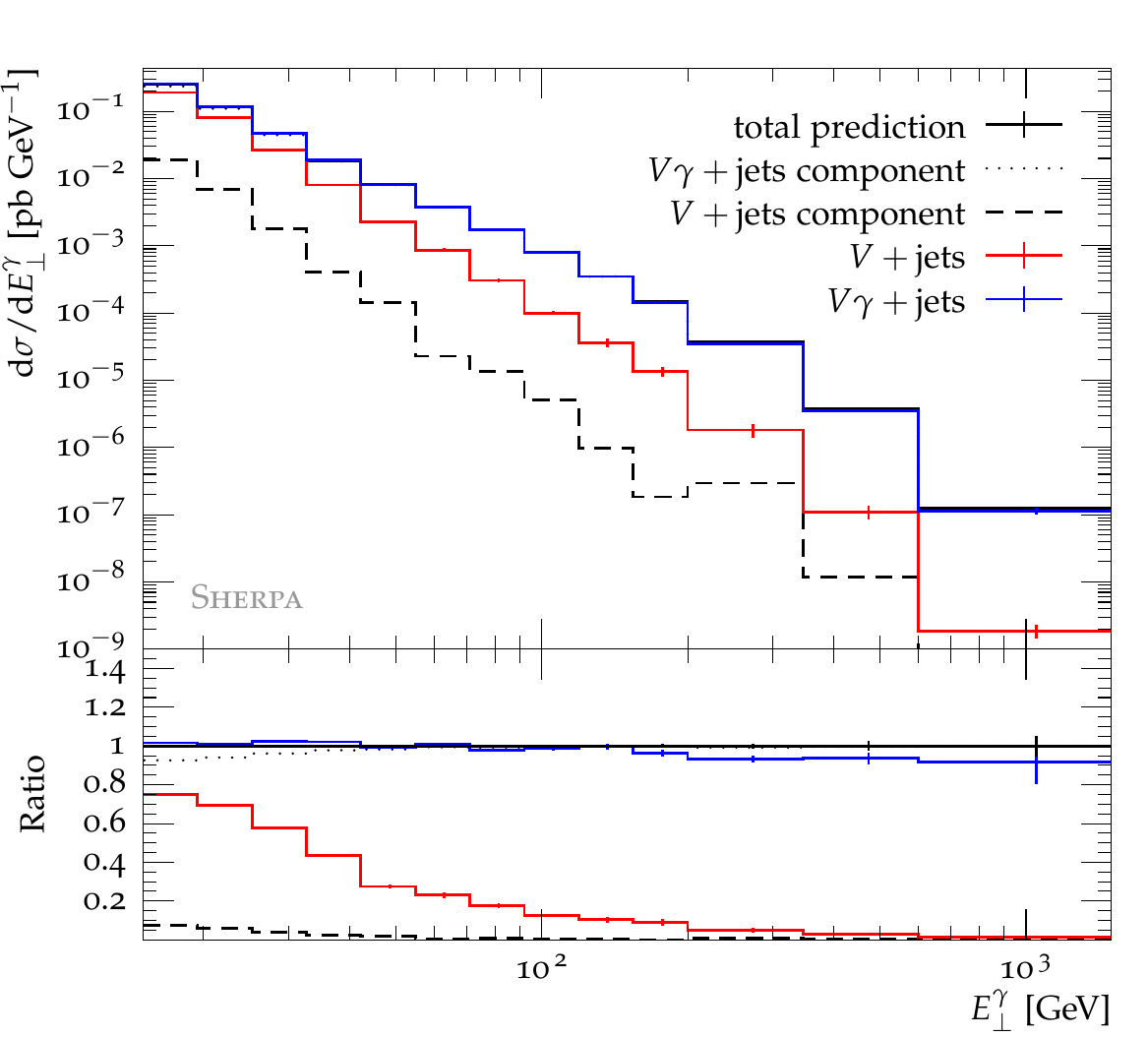}}
  \subfloat[inclusive jet multiplicities]{\includegraphics[width=0.5\linewidth]{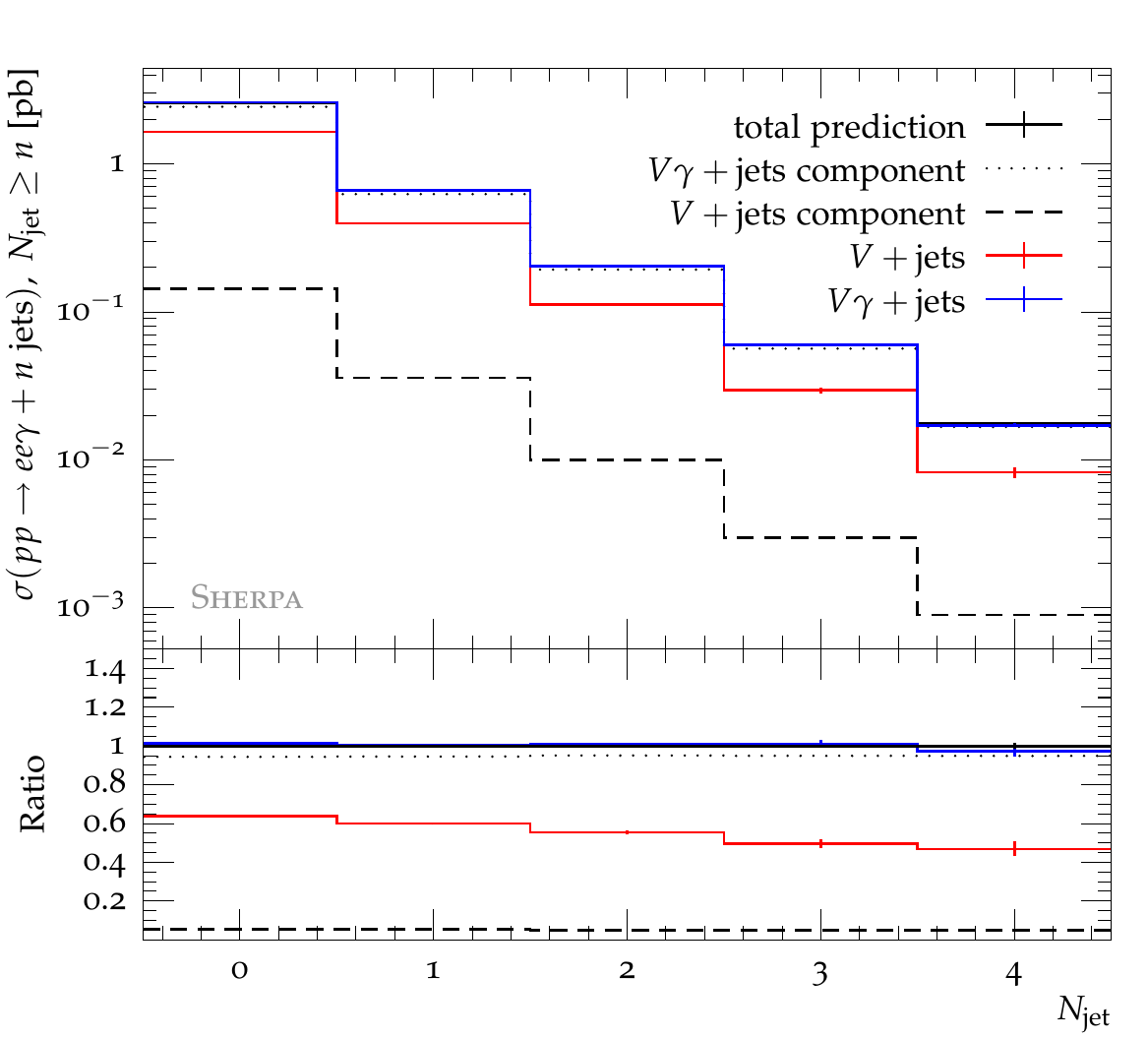}}
    \caption{Comparison of the overlap removal procedure (OR) with a pure $Z\gamma$ and a pure $Z$ sample in the $Z\gamma$ control region (region \RM{1}).}
   \label{fig:fsr_zgamma_res} 
\end{figure*}

\begin{figure*}[p]
  \centering 
  \subfloat[$Z$ mass, constructed of dressed leptons]{\includegraphics[width=0.5\linewidth]{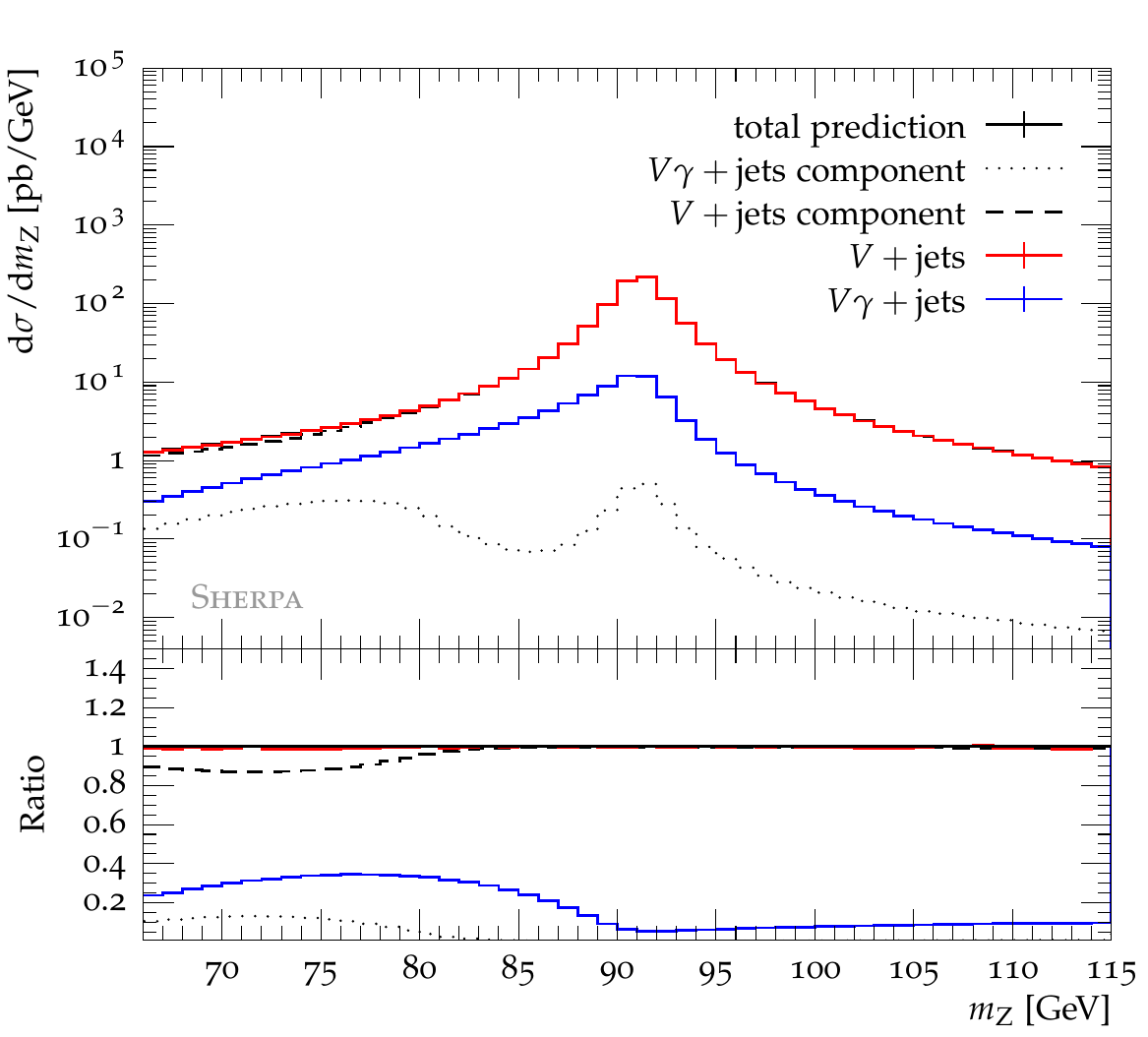}}
  \subfloat[$p_\perp$ of $Z$ boson]{\includegraphics[width=0.5\linewidth]{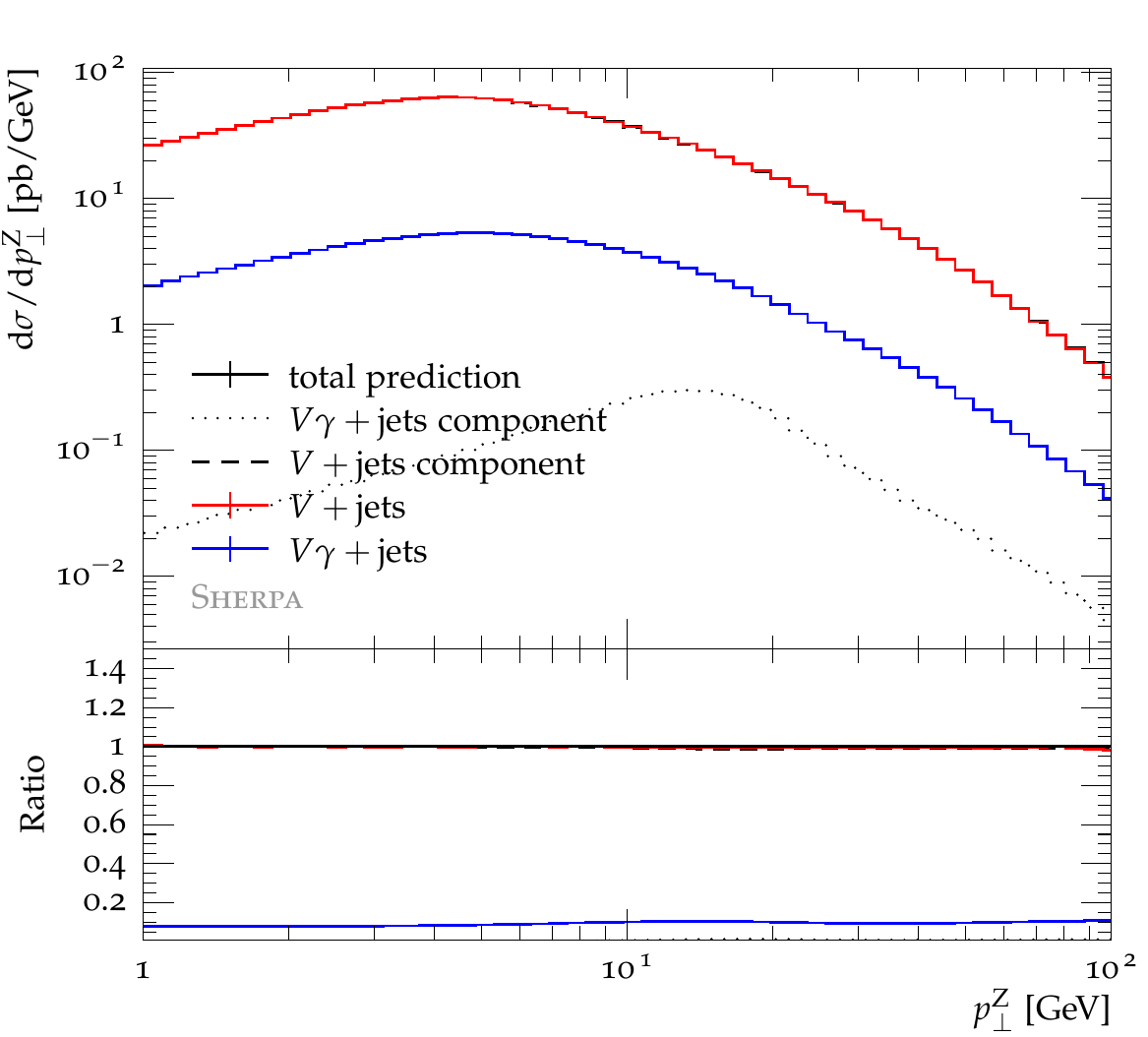}}
    \caption{Comparison of the overlap removal procedure (OR) with a pure $Z\gamma$ and a pure $Z$ sample in the inclusive $Z$ control region (region \RM{2}).}
   \label{fig:fsr_z_res} 
\end{figure*}

\begin{figure*}[p]
  \centering 
  \subfloat[$E_\perp^\gamma$ spectrum, $0.05<\DELTA R_{\gamma e^\pm} < 3$]{\includegraphics[width=0.5\linewidth]{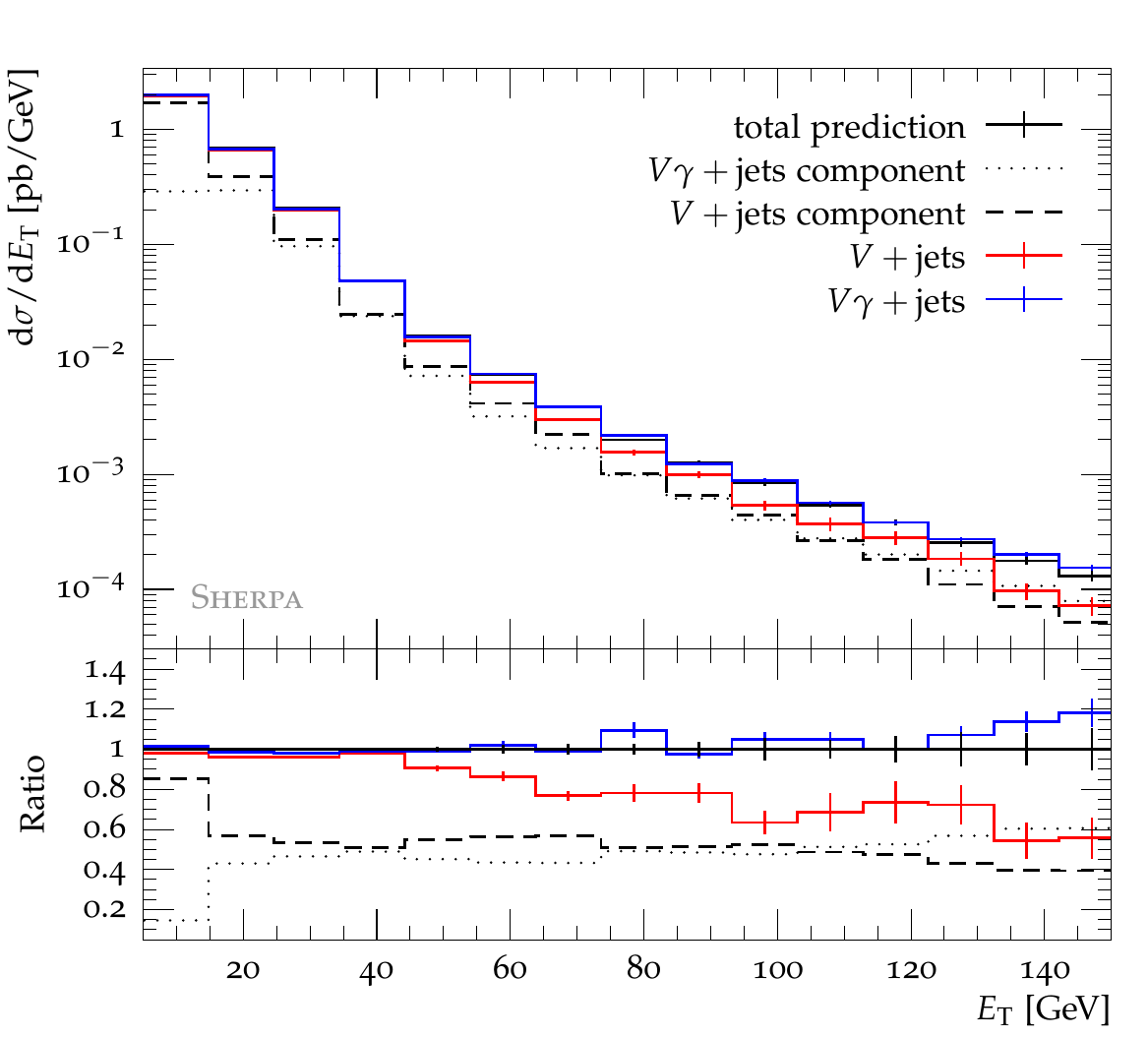}}

  \subfloat[$E_\perp^\gamma$ spectrum, $0.05<\DELTA R_{\gamma e^\pm} < 0.5$]{\includegraphics[width=0.5\linewidth]{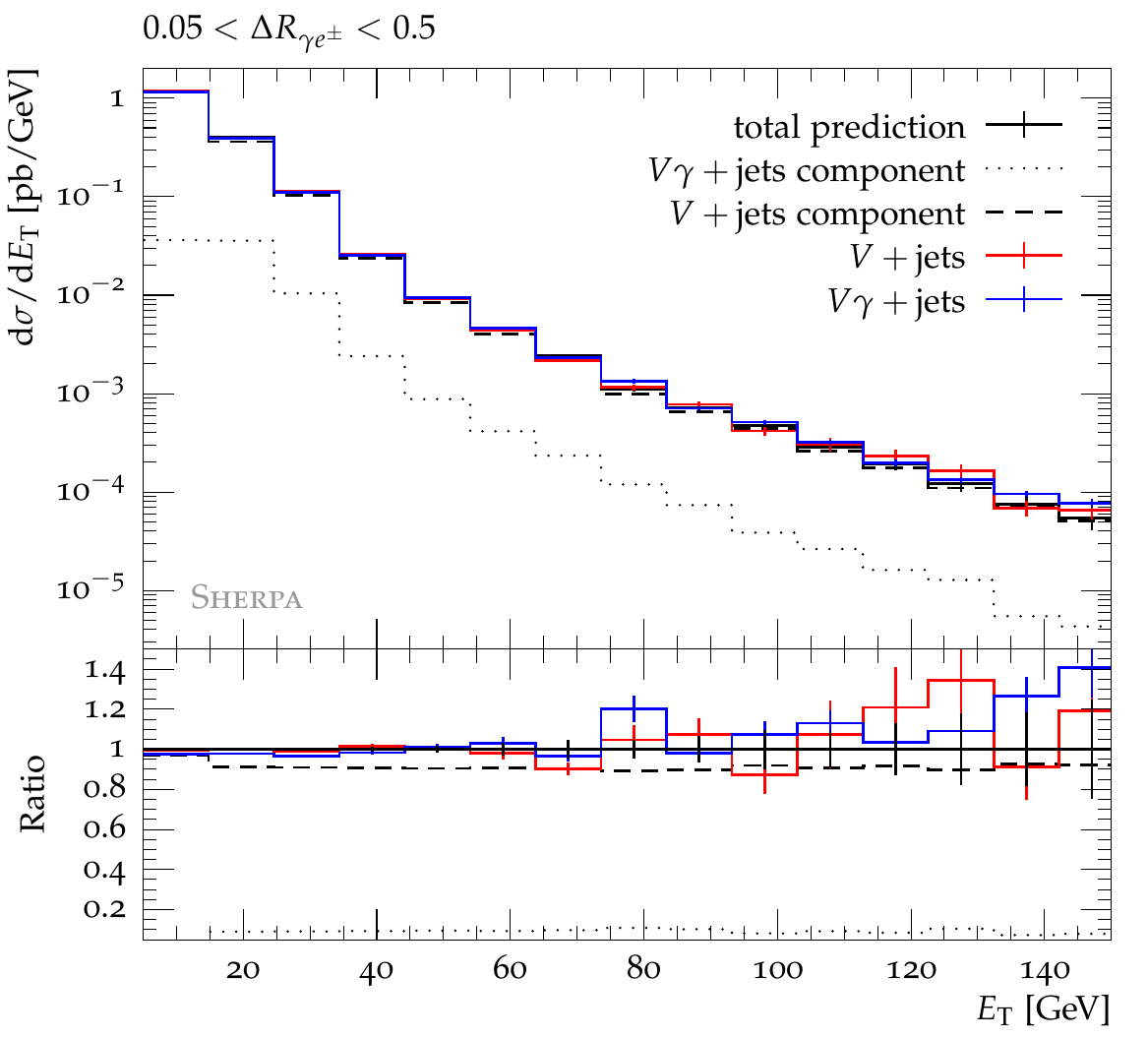}}
  \subfloat[$E_\perp^\gamma$ spectrum, $0.5<\DELTA R_{\gamma e^\pm} < 3$]{\includegraphics[width=0.5\linewidth]{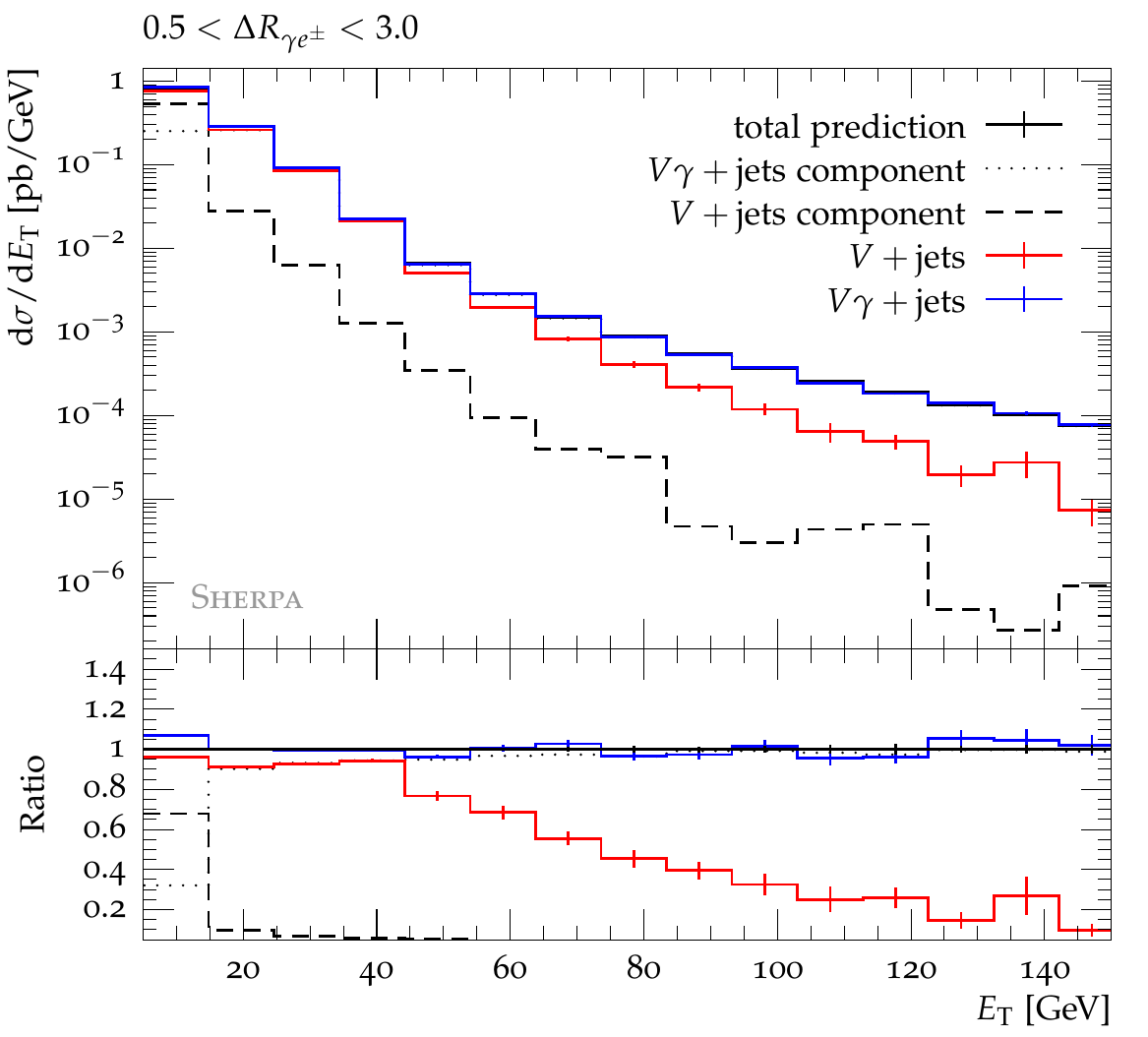}}
    \caption{Comparison of the overlap removal procedure (OR) with a pure $Z\gamma$ and a pure $Z$ sample in a final state radiation dominated test region (region \RM{3}). $\DELTA R_{\gamma e^\pm}$ refers to the angular distance between the photon and the closest electron / positron. }
   \label{fig:fsr_fsr_res_et} 
\end{figure*}

\begin{figure*}[p]
  \centering 
  \subfloat[azimuthal distance between the photon and its closest lepton]{\includegraphics[width=0.5\linewidth]{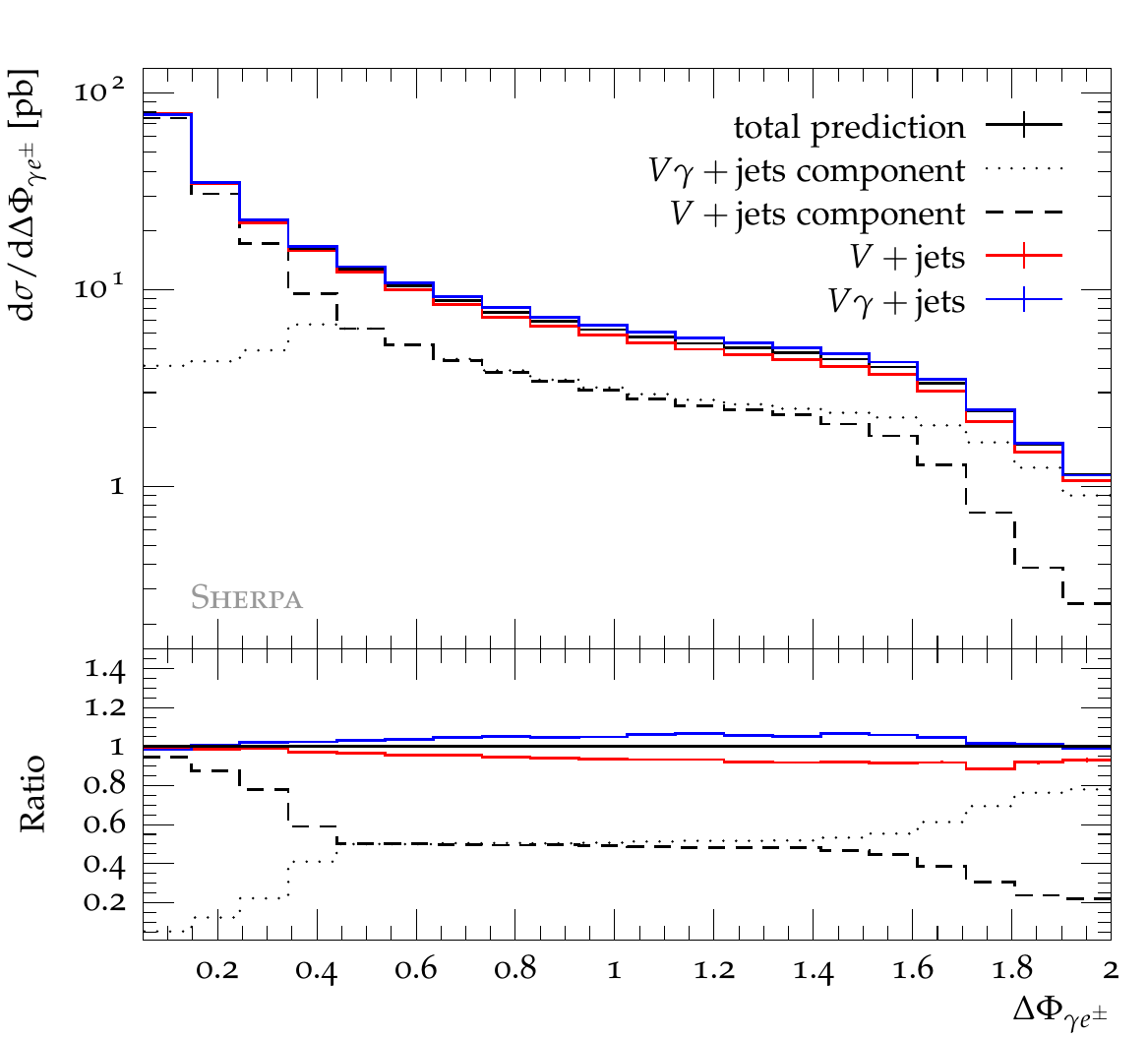}}
  \subfloat[invariant mass of both leptons and the photon]{\includegraphics[width=0.5\linewidth]{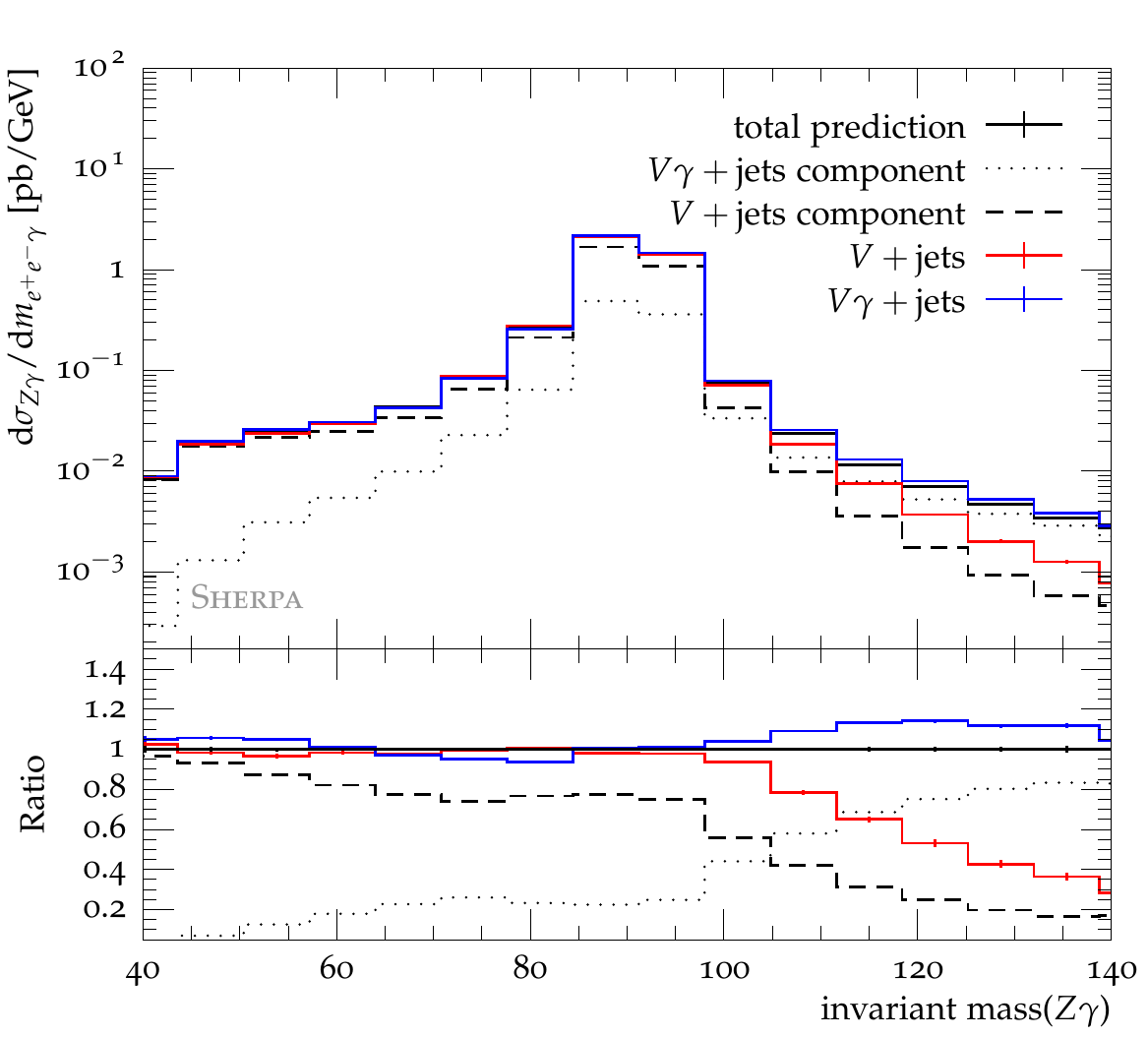}}
    \caption{Comparison  of the overlap removal procedure (OR) with a pure $Z\gamma$ and a pure $Z$ sample for further observables in a final state radiation dominated test region (region \RM{3}). }
   \label{fig:fsr_fsr_res_other} 
\end{figure*}

The validation of the overlap removal algorithm proceeds with analyses in three different phase space regions. 
Both the $Z\gamma$ and $Z$ prediction should not be altered by the overlap removal in their regions of validity.
The first condition is checked using the $Z\gamma$ analysis introduced in Section~\ref{subsec:results_13_dir}. It covers the explicit $Z\gamma$ phase space and defines region \RM{1}.
By contrast, the $Z$ phase space is probed by the default inclusive $Z$ analysis provided by \textsc{Rivet}. Here, a reconstructed $Z$ boson with an invariant mass between 65 and 115\gev\ is required. The leptons are dressed with all photons having an angular distance of 0.2 or smaller. These cuts define the phase space region \RM{2}. 

In addition, a special region \RM{3} is defined where final state radiation contributions via YFS are supposed to give similar contributions as the direct production from matrix elements. Having such a region it is directly possible to study the interplay between both components of the overlap removal and compare their sum with a pure YFS or direct sample. 

A region dominated by final state radiation is defined by requiring the lepton pair to have an invariant mass below the $Z$ peak, $30\gev < m^{ll} < 87\gev$. 
An event is accepted, if both leading leptons have electron flavour but opposite charge. 
As photon candidate the leading photon ($p_\perp^\gamma > 5\gev $) is chosen, it has to be isolated from the selected leptons by requiring $\DELTA R^{\gamma, e^\pm} > 0.05 $. 
In addition, the total energy of all remaining particles (excluding the selected leptons) in a cone with $\DELTA R<0.4$ around the photon axis has to be less than $0.5 \cdot E_\perp^\gamma$. 
Details of all analyses are summarized in Table~\ref{tab:fsrcuts}. All cuts and observables are implemented as a user module using the \textsc{Rivet} framework.

A comparison is performed using four sets of separately generated event samples; pure $Z\gamma$, pure $Z$, sliced $Z$ and sliced $Z\gamma$.  The latter two are summed up to give the total prediction after overlap removal. YFS is set active for all samples. 

For event generation, the matrix element level cuts are selected to be more inclusive than the analysis cuts. For the generation of the sliced direct part, the matrix level cuts are additionally chosen to be more inclusive than the phase space slicing parameters. All events are generated with $Q_\mathrm{CUT}=30$\gev\ and up to 3 jets at leading-order accuracy\footnote{MEPS@LO is chosen simply for performance reasons. This implies no limitation as long as slicing cuts are IR save since the introduced slicing algorithm is based solely on kinematics.}.  The slicing parameters which are used for this test are summarized in Table~\ref{tab:slicing_param}.

In Figure~\ref{fig:fsr_zgamma_res}, the inclusive jet multiplicity and the
inclusive $E_\perp^\gamma$ spectrum for the $Z\gamma$ phase space (region \RM{1}) are shown.
 In these and all further
plots the inclusive $Z$, the direct $Z\gamma$ and the summed overlap-removed predictions are
shown, together with the corresponding components in the overlap removal. For better readability the statistical uncertainties of the latter have been omitted. 

Here, the overlap removal result is in very good agreement with the pure $Z\gamma$ prediction in both plots. The dominating contribution is the $Z\gamma$ component, giving about 90\% of the cross section for low $p_\perp$ and almost 100\% if $p_\perp$ exceeds 80\gev. 

By contrast, the inclusive $Z$ phase space (region \RM{2}) is dominated by the $Z$ component. The $Z$ mass and $Z$ $p_\perp$ distributions are shown in Figure~\ref{fig:fsr_z_res}. The overlap removal result is in excellent agreement with the pure $Z$ prediction. The only region of phase space where the direct component of the overlap removal is sizeable is the low mass region of less than 85\gev, there the direct component gives around 10\% of the cross section.

Despite the good agreement between overlap removal and the respective reference, one might want to construct the overlap removal to require only the $Z\gamma$ component when looking at $Z\gamma$ analyses. This would require to take into account the analysis cuts for slicing the phase space and is thus not possible in a generic sample. An overlap-removed contribution with inclusive $Z$ production allows more flexibility as needed in general purpose experiments.

Finally, in Figure~\ref{fig:fsr_fsr_res_et} and~\ref{fig:fsr_fsr_res_other} some observables of the  FSR dominated phase space (region \RM{3}) are shown. 
In this region neither the pure $Z$ nor the pure $Z\gamma$ predictions are guaranteed to give an accurate result. 
The former one includes only final state radiation and will therefore miss contributions especially in the high $E_\perp$ region. 
By contrast, the latter one includes all contributions at a fixed order but leaves the perturbative region if the photon is soft or very close to the lepton. Thus, both these predictions can only be interpreted as lower and upper bounds for the overlap removal in the context of this validation. 

In Figure~\ref{fig:fsr_fsr_res_et}, different $E_\perp^\gamma$ spectra are shown. The first subplot covers the whole phase space while the two remaining ones cover only the regions where the photon is either very close to ($0.05<\DELTA R<0.5$) or separated from ($0.5 < \DELTA R<3$) the closest lepton. While in the inclusive plot both components of the overlap removal procedure give a very similar contribution if $E_\perp^\gamma$ exceeds the slicing cut, the two remaining plots reveal the nature of the overlap removal procedure much better. Whereas the low $\DELTA R$ region is entirely dominated by the $Z$ component, the high $\DELTA R$ region is dominated by the $Z\gamma$ component as soon as the cut off is exceeded.

Figure~\ref{fig:fsr_fsr_res_other} shows the azimuthal angle between the photon and its closest lepton and the invariant mass of the photon and both leptons. Both observables show an interesting behaviour. As expected, the cross section of the combined overlap removal result always interpolates between the cross section of the pure $Z\gamma$ and the pure $Z$ sample. When looking at the azimuthal distance, the $Z$ component dominates at lower and the $Z\gamma$ component at higher values, while both components are equal in a large region of phase space. The invariant mass spectrum is dominated at lower values ($<105$\gev) by the $Z$ component and at higher values by $Z\gamma$.

\begin{table}[]
\centering
\begin{tabular}{|c|c|} \hline
\multicolumn{2}{|c|}{region \RM{1}, $Z\gamma$ control region} \\ \hline
Lepton & $p_\perp>25$\gev, $\abs{\eta}<2.5$  \\
Jet    & $E_\perp>30$\gev, $\abs{\eta}<4.4$, $\DELTA R(\mathrm{jet}$, $e/\gamma ) > 0.3$                   \\
Boson  & $M_{e^+, e^-}>40$\gev          \\
Photon & $E_\perp>15$\gev, $\abs{\eta}<2.5$ \\
Isolation & $\DELTA R(\gamma, e^\pm)>0.4$, $\epsilon_{\mathrm{iso}}=0.5$  \\
\hline
\multicolumn{2}{|c|}{region \RM{2}, inclusive $Z$ control region} \\ \hline
Leptons & $p_\perp>25$\gev, $\abs{\eta}<3.5$, opposite charge  \\
$Z$  & $65\gev < M^{ll} <115$\gev                                                                                \\
\hline
\multicolumn{2}{|c|}{region \RM{3}, overlap removal test region} \\ \hline
Leptons & $p_\perp>15$\gev, $\abs{\eta}<2.5$, opposite charge  \\
$Z$  & $30\gev < M^{ll} <87$\gev                                                                                \\
Photon & $E_\perp>5$\gev, $\abs{\eta}<2.5$ \\
\hline
\end{tabular}
\caption{This table summarizes all cuts which define the phase space region which is used for testing the introduced overlap removal procedure.}
\label{tab:fsrcuts}
\end{table}

\begin{table}[]
\centering
\begin{tabular}{|c|c|} \hline
hardest photon &  $p_\perp>10$\gev \\
photon lepton isolation  & $ \DELTA R (\mathrm{leptons}, \gamma) > 0.4 $ \\
photon hadron isolation  & $R=0.4$, $n=1$, $\epsilon=0.5$ \\ 
\hline
\end{tabular}
\caption{Slicing parameters which are used for validation of the overlap removal procedure. These cuts are applied according to the procedure defined in Section \ref{subec:or_imp}}
\label{tab:slicing_param}
\end{table}

\FloatBarrier

\section{Conclusions}

Precise Standard Model predictions for $Z\gamma$+jets production
are crucial for the search for new particles or anomalous couplings in
measurements of this final state at the LHC.

With the presented simulation within the \textsc{Sherpa} framework using
the MEPS@NLO algorithm we provide a simulation which is at the same time
precise and realistic: NLO QCD corrections for the $Z\gamma$ and $Z\gamma$+jet
processes are included and reduce the uncertainties in relevant observables
significantly. At the same time, the matching and merging with the parton
shower allows a realistic simulation of the full final state at the hadron
level, and the inclusion of all off-shell effects allows to place realistic
experimental cuts on the prompt leptons without approximations.

Comparing to data from experimental measurements at $\sqrt{s}=8\tev$ we find
very good agreement. On that basis we make predictions at $\sqrt{s}=13\tev$
and identify the dominant theoretical uncertainties and the phase space
regions affected by them.

To further the application of these precise $Z\gamma$+jets predictions
in experiments we also demonstrate how they can be combined with event
generator predictions for $Z$+jets including final-state photon radiation.
As a validation we introduce a number of cross checks based on different
phase space regions which can be repeated for any specific application of
such samples in the experiments.

\begin{acknowledgements}
  We are grateful to our colleagues in the \textsc{Atlas} and \textsc{Sherpa} collaborations for
  useful discussions and support, in particular to Marek~Sch\"onherr for his comments on the
  manuscript. We thank the OpenLoops authors for providing
  the necessary virtual matrix elements.
  This research was supported by the German Research Foundation (DFG) under
  grant No.\ SI 2009/1-1.
\end{acknowledgements}

\newpage

\bibliographystyle{bib/amsunsrt_modpp}       
\bibliography{bib/journal}

\ifx\mcitethebibliography\mciteundefinedmacro
\PackageError{amsunsrt_mod.bst}{mciteplus.sty has not been loaded}
{This bibstyle requires the use of the mciteplus package.}\fi
\begin{mcitethebibliography}{10}

\bibitem{Achard:2004ds}
P.~Achard et~al., L3, \emph{{Study of the $e^{+} e^{-} \to Z \gamma$ process at
  LEP and limits on triple neutral-gauge-boson couplings}}, Phys. Lett.
  \textbf{B597} (2004),
  \href{http://inspirehep.net/search?p=hep-ex/0407012}{119--130},
  [\href{http://arXiv.org/pdf/hep-ex/0407012}{{\texttt{ arXiv:hep-ex/0407012}}}
  [hep-ex]]%
\relax\mciteBstWouldAddEndPuncttrue
\mciteSetBstMidEndSepPunct{\mcitedefaultmidpunct}
{\mcitedefaultendpunct}{\mcitedefaultseppunct}\relax
\EndOfBibitem
\bibitem{Abdallah:2007ae}
J.~Abdallah et~al., DELPHI, \emph{{Study of triple-gauge-boson couplings $ZZZ$,
  $ZZ\gamma$ and $Z\gamma\gamma$ at LEP}}, Eur. Phys. J. \textbf{C51} (2007),
  \href{http://inspirehep.net/search?p=0706.2741}{525--542},
  [\href{http://arXiv.org/pdf/0706.2741}{{\texttt{ arXiv:0706.2741}}}
  [hep-ex]]%
\relax\mciteBstWouldAddEndPuncttrue
\mciteSetBstMidEndSepPunct{\mcitedefaultmidpunct}
{\mcitedefaultendpunct}{\mcitedefaultseppunct}\relax
\EndOfBibitem
\bibitem{Abbiendi:2000cu}
G.~Abbiendi et~al., OPAL, \emph{{Search for trilinear neutral gauge boson
  couplings in $Z^-$ gamma production at $\sqrt{s}$ = 189 GeV at LEP}}, Eur.
  Phys. J. \textbf{C17} (2000),
  \href{http://inspirehep.net/search?p=hep-ex/0007016}{553--566},
  [\href{http://arXiv.org/pdf/hep-ex/0007016}{{\texttt{ arXiv:hep-ex/0007016}}}
  [hep-ex]]%
\relax\mciteBstWouldAddEndPuncttrue
\mciteSetBstMidEndSepPunct{\mcitedefaultmidpunct}
{\mcitedefaultendpunct}{\mcitedefaultseppunct}\relax
\EndOfBibitem
\bibitem{Abbiendi:2004bf}
G.~Abbiendi et~al., OPAL, \emph{{Constraints on anomalous quartic gauge boson
  couplings from $\nu \bar{\nu} \gamma \gamma$ and $q \bar{q} \gamma \gamma$
  events at LEP-2}}, Phys. Rev. \textbf{D70} (2004),
  \href{http://inspirehep.net/search?p=hep-ex/0402021}{032005},
  [\href{http://arXiv.org/pdf/hep-ex/0402021}{{\texttt{ arXiv:hep-ex/0402021}}}
  [hep-ex]]%
\relax\mciteBstWouldAddEndPuncttrue
\mciteSetBstMidEndSepPunct{\mcitedefaultmidpunct}
{\mcitedefaultendpunct}{\mcitedefaultseppunct}\relax
\EndOfBibitem
\bibitem{Abazov:2009cj}
V.~M. Abazov et~al., D0, \emph{{Measurement of the $Z \gamma \to \nu \bar\nu
  \gamma$ cross section and limits on anomalous $Z Z \gamma$ and $Z \gamma
  gamma$ couplings in p anti-p collisions at $\sqrt{s}=1.96$ TeV}}, Phys. Rev.
  Lett. \textbf{102} (2009),
  \href{http://inspirehep.net/search?p=0902.2157}{201802},
  [\href{http://arXiv.org/pdf/0902.2157}{{\texttt{ arXiv:0902.2157}}}
  [hep-ex]]%
\relax\mciteBstWouldAddEndPuncttrue
\mciteSetBstMidEndSepPunct{\mcitedefaultmidpunct}
{\mcitedefaultendpunct}{\mcitedefaultseppunct}\relax
\EndOfBibitem
\bibitem{Abazov:2011qp}
V.~M. Abazov et~al., D0, \emph{{$Z\gamma$ production and limits on anomalous
  $ZZ\gamma$ and $Z\gamma\gamma$ couplings in $p\bar{p}$ collisions at
  $\sqrt{s}=1.96$ TeV}}, Phys. Rev. \textbf{D85} (2012),
  \href{http://inspirehep.net/search?p=1111.3684}{052001},
  [\href{http://arXiv.org/pdf/1111.3684}{{\texttt{ arXiv:1111.3684}}}
  [hep-ex]]%
\relax\mciteBstWouldAddEndPuncttrue
\mciteSetBstMidEndSepPunct{\mcitedefaultmidpunct}
{\mcitedefaultendpunct}{\mcitedefaultseppunct}\relax
\EndOfBibitem
\bibitem{Aaltonen:2011zc}
T.~Aaltonen et~al., CDF, \emph{{Limits on Anomalous Trilinear Gauge Couplings
  in $Z\gamma$ Events from $p\bar{p}$ Collisions at $\sqrt{s} = 1.96$ TeV}},
  Phys. Rev. Lett. \textbf{107} (2011),
  \href{http://inspirehep.net/search?p=1103.2990}{051802},
  [\href{http://arXiv.org/pdf/1103.2990}{{\texttt{ arXiv:1103.2990}}}
  [hep-ex]]%
\relax\mciteBstWouldAddEndPuncttrue
\mciteSetBstMidEndSepPunct{\mcitedefaultmidpunct}
{\mcitedefaultendpunct}{\mcitedefaultseppunct}\relax
\EndOfBibitem
\bibitem{Aad:2013izg}
G.~Aad et~al., ATLAS, \emph{{Measurements of $W \gamma$ and $Z \gamma$
  production in $pp$ collisions at $\sqrt{s}=7$ TeV with the ATLAS detector at
  the LHC}}, Phys. Rev. \textbf{D87} (2013), no.~11,
  \href{http://inspirehep.net/search?p=1302.1283}{112003},
  [\href{http://arXiv.org/pdf/1302.1283}{{\texttt{ arXiv:1302.1283}}}
  [hep-ex]], [Erratum: Phys. Rev.D91,no.11,119901(2015)]%
\relax\mciteBstWouldAddEndPuncttrue
\mciteSetBstMidEndSepPunct{\mcitedefaultmidpunct}
{\mcitedefaultendpunct}{\mcitedefaultseppunct}\relax
\EndOfBibitem
\bibitem{Aad:2016sau}
G.~Aad et~al., ATLAS, \emph{{Measurements of $Z\gamma$ and $Z\gamma\gamma$
  production in $pp$ collisions at $\sqrt{s}=$ 8 TeV with the ATLAS detector}},
  Phys. Rev. \textbf{D93} (2016), no.~11,
  \href{http://inspirehep.net/search?p=1604.05232}{112002},
  [\href{http://arXiv.org/pdf/1604.05232}{{\texttt{ arXiv:1604.05232}}}
  [hep-ex]]%
\relax\mciteBstWouldAddEndPuncttrue
\mciteSetBstMidEndSepPunct{\mcitedefaultmidpunct}
{\mcitedefaultendpunct}{\mcitedefaultseppunct}\relax
\EndOfBibitem
\bibitem{Aaboud:2017pds}
M.~Aaboud et~al., ATLAS, \emph{{Studies of $Z\gamma$ production in association
  with a high-mass dijet system in $pp$ collisions at $\sqrt{s}=$ 8 TeV with
  the ATLAS detector}}, JHEP \textbf{07} (2017),
  \href{http://inspirehep.net/search?p=1705.01966}{107},
  [\href{http://arXiv.org/pdf/1705.01966}{{\texttt{ arXiv:1705.01966}}}
  [hep-ex]]%
\relax\mciteBstWouldAddEndPuncttrue
\mciteSetBstMidEndSepPunct{\mcitedefaultmidpunct}
{\mcitedefaultendpunct}{\mcitedefaultseppunct}\relax
\EndOfBibitem
\bibitem{Chatrchyan:2013nda}
S.~Chatrchyan et~al., CMS, \emph{{Measurement of the production cross section
  for $Z\gamma \to \nu\bar{\nu}\gamma$ in pp collisions at $\sqrt{s} =$ 7 TeV
  and limits on $ZZ\gamma$ and $Z\gamma\gamma$ triple gauge boson couplings}},
  JHEP \textbf{10} (2013),
  \href{http://inspirehep.net/search?p=1309.1117}{164},
  [\href{http://arXiv.org/pdf/1309.1117}{{\texttt{ arXiv:1309.1117}}}
  [hep-ex]]%
\relax\mciteBstWouldAddEndPuncttrue
\mciteSetBstMidEndSepPunct{\mcitedefaultmidpunct}
{\mcitedefaultendpunct}{\mcitedefaultseppunct}\relax
\EndOfBibitem
\bibitem{Chatrchyan:2013fya}
S.~Chatrchyan et~al., CMS, \emph{{Measurement of the $W\gamma$ and $Z\gamma$
  inclusive cross sections in $pp$ collisions at $\sqrt s=7$ TeV and limits on
  anomalous triple gauge boson couplings}}, Phys. Rev. \textbf{D89} (2014),
  no.~9, \href{http://inspirehep.net/search?p=1308.6832}{092005},
  [\href{http://arXiv.org/pdf/1308.6832}{{\texttt{ arXiv:1308.6832}}}
  [hep-ex]]%
\relax\mciteBstWouldAddEndPuncttrue
\mciteSetBstMidEndSepPunct{\mcitedefaultmidpunct}
{\mcitedefaultendpunct}{\mcitedefaultseppunct}\relax
\EndOfBibitem
\bibitem{Khachatryan:2015kea}
V.~Khachatryan et~al., CMS, \emph{{Measurement of the $\mathrm{Z}\gamma$
  Production Cross Section in pp Collisions at 8 TeV and Search for Anomalous
  Triple Gauge Boson Couplings}}, JHEP \textbf{04} (2015),
  \href{http://inspirehep.net/search?p=1502.05664}{164},
  [\href{http://arXiv.org/pdf/1502.05664}{{\texttt{ arXiv:1502.05664}}}
  [hep-ex]]%
\relax\mciteBstWouldAddEndPuncttrue
\mciteSetBstMidEndSepPunct{\mcitedefaultmidpunct}
{\mcitedefaultendpunct}{\mcitedefaultseppunct}\relax
\EndOfBibitem
\bibitem{Khachatryan:2016yro}
V.~Khachatryan et~al., CMS, \emph{{Measurement of the $ \mathrm{ Z } \gamma
  \rightarrow \nu \bar{\nu} \gamma$ production cross section in pp collisions
  at $\sqrt{s}=$ 8 TeV and limits on anomalous $ \mathrm{ ZZ } \gamma$ and $
  \mathrm{Z} \gamma \gamma$ trilinear gauge boson couplings}}, Phys. Lett.
  \textbf{B760} (2016),
  \href{http://inspirehep.net/search?p=1602.07152}{448--468},
  [\href{http://arXiv.org/pdf/1602.07152}{{\texttt{ arXiv:1602.07152}}}
  [hep-ex]]%
\relax\mciteBstWouldAddEndPuncttrue
\mciteSetBstMidEndSepPunct{\mcitedefaultmidpunct}
{\mcitedefaultendpunct}{\mcitedefaultseppunct}\relax
\EndOfBibitem
\bibitem{Djouadi:1996yq}
A.~Djouadi, V.~Driesen, W.~Hollik and A.~Kraft, \emph{{The Higgs photon - Z
  boson coupling revisited}}, Eur.Phys.J. \textbf{C1} (1998),
  \href{http://inspirehep.net/record/428253}{163--175},
  [\href{http://arXiv.org/pdf/hep-ph/9701342}{{\texttt{ arXiv:hep-ph/9701342}}}
  [hep-ph]]%
\relax\mciteBstWouldAddEndPuncttrue
\mciteSetBstMidEndSepPunct{\mcitedefaultmidpunct}
{\mcitedefaultendpunct}{\mcitedefaultseppunct}\relax
\EndOfBibitem
\bibitem{Aaboud:2016trl}
M.~Aaboud et~al., ATLAS, \emph{{Search for heavy resonances decaying to a $Z$
  boson and a photon in $pp$ collisions at $\sqrt{s}=13$ TeV with the ATLAS
  detector}}, Phys. Lett. \textbf{B764} (2017),
  \href{http://inspirehep.net/search?p=1607.06363}{11--30},
  [\href{http://arXiv.org/pdf/1607.06363}{{\texttt{ arXiv:1607.06363}}}
  [hep-ex]]%
\relax\mciteBstWouldAddEndPuncttrue
\mciteSetBstMidEndSepPunct{\mcitedefaultmidpunct}
{\mcitedefaultendpunct}{\mcitedefaultseppunct}\relax
\EndOfBibitem
\bibitem{Khachatryan:2016odk}
V.~Khachatryan et~al., CMS, \emph{{Search for high-mass Z$\gamma$ resonances in
  $\mathrm{ e }^{+}\mathrm{ e }^{-}\gamma $ and $ \mu^{+}\mu^{-}\gamma$ final
  states in proton-proton collisions at $\sqrt{s} =$ 8 and 13 TeV}}, JHEP
  \textbf{01} (2017), \href{http://inspirehep.net/search?p=1610.02960}{076},
  [\href{http://arXiv.org/pdf/1610.02960}{{\texttt{ arXiv:1610.02960}}}
  [hep-ex]]%
\relax\mciteBstWouldAddEndPuncttrue
\mciteSetBstMidEndSepPunct{\mcitedefaultmidpunct}
{\mcitedefaultendpunct}{\mcitedefaultseppunct}\relax
\EndOfBibitem
\bibitem{Aaboud:2016uro}
M.~Aaboud et~al., ATLAS, \emph{{Search for new phenomena in events with a
  photon and missing transverse momentum in $pp$ collisions at $\sqrt{s}=13$
  TeV with the ATLAS detector}}, JHEP \textbf{06} (2016),
  \href{http://inspirehep.net/search?p=1604.01306}{059},
  [\href{http://arXiv.org/pdf/1604.01306}{{\texttt{ arXiv:1604.01306}}}
  [hep-ex]]%
\relax\mciteBstWouldAddEndPuncttrue
\mciteSetBstMidEndSepPunct{\mcitedefaultmidpunct}
{\mcitedefaultendpunct}{\mcitedefaultseppunct}\relax
\EndOfBibitem
\bibitem{Aaboud:2017dor}
M.~Aaboud et~al., ATLAS, \emph{{Search for dark matter at $\sqrt{s}=13$ TeV in
  final states containing an energetic photon and large missing transverse
  momentum with the ATLAS detector}}, Eur. Phys. J. \textbf{C77} (2017), no.~6,
  \href{http://inspirehep.net/search?p=1704.03848}{393},
  [\href{http://arXiv.org/pdf/1704.03848}{{\texttt{ arXiv:1704.03848}}}
  [hep-ex]]%
\relax\mciteBstWouldAddEndPuncttrue
\mciteSetBstMidEndSepPunct{\mcitedefaultmidpunct}
{\mcitedefaultendpunct}{\mcitedefaultseppunct}\relax
\EndOfBibitem
\bibitem{Khachatryan:2016ojf}
V.~Khachatryan et~al., CMS, \emph{{Search for supersymmetry in events with
  photons and missing transverse energy in pp collisions at 13 TeV}}, Phys.
  Lett. \textbf{B769} (2017),
  \href{http://inspirehep.net/search?p=1611.06604}{391--412},
  [\href{http://arXiv.org/pdf/1611.06604}{{\texttt{ arXiv:1611.06604}}}
  [hep-ex]]%
\relax\mciteBstWouldAddEndPuncttrue
\mciteSetBstMidEndSepPunct{\mcitedefaultmidpunct}
{\mcitedefaultendpunct}{\mcitedefaultseppunct}\relax
\EndOfBibitem
\bibitem{Sirunyan:2017ewk}
\href{http://inspirehep.net/search?p=1706.03794}{A.~M. Sirunyan et~al.}, CMS,
  \emph{{Search for new physics in the monophoton final state in proton-proton
  collisions at sqrt(s) = 13 TeV}},
  \href{http://arXiv.org/pdf/1706.03794}{{\texttt{ arXiv:1706.03794}}}
  [hep-ex]%
\relax\mciteBstWouldAddEndPuncttrue
\mciteSetBstMidEndSepPunct{\mcitedefaultmidpunct}
{\mcitedefaultendpunct}{\mcitedefaultseppunct}\relax
\EndOfBibitem
\bibitem{Renard:1981es}
F.~M. Renard, \emph{{Tests of Neutral Gauge Boson Selfcouplings With $e^+ e^-
  \to \gamma Z$}}, Nucl. Phys. \textbf{B196} (1982),
  \href{http://inspirehep.net/search?j=Nucl%20Phys,B196,93}{93--108},
  CERN-TH-3185%
\relax\mciteBstWouldAddEndPuncttrue
\mciteSetBstMidEndSepPunct{\mcitedefaultmidpunct}
{\mcitedefaultendpunct}{\mcitedefaultseppunct}\relax
\EndOfBibitem
\bibitem{Ohnemus:1992jn}
J.~Ohnemus, \emph{{Order $\alpha^- s$ calculations of hadronic $W^\pm \gamma$
  and $Z \gamma$ production}}, Phys. Rev. \textbf{D47} (1993),
  \href{http://inspirehep.net/search?j=Phys%20Rev,D47,940}{940--955},
  DTP-92-54%
\relax\mciteBstWouldAddEndPuncttrue
\mciteSetBstMidEndSepPunct{\mcitedefaultmidpunct}
{\mcitedefaultendpunct}{\mcitedefaultseppunct}\relax
\EndOfBibitem
\bibitem{Ohnemus:1994qp}
J.~Ohnemus, \emph{{Hadronic $Z \gamma$ production with QCD corrections and
  leptonic decays}}, Phys. Rev. \textbf{D51} (1995),
  \href{http://inspirehep.net/search?p=hep-ph/9407370}{1068--1076},
  [\href{http://arXiv.org/pdf/hep-ph/9407370}{{\texttt{ arXiv:hep-ph/9407370}}}
  [hep-ph]]%
\relax\mciteBstWouldAddEndPuncttrue
\mciteSetBstMidEndSepPunct{\mcitedefaultmidpunct}
{\mcitedefaultendpunct}{\mcitedefaultseppunct}\relax
\EndOfBibitem
\bibitem{Baur:1997kz}
U.~Baur, T.~Han and J.~Ohnemus, \emph{{QCD corrections and anomalous couplings
  in $Z \gamma$ production at hadron colliders}}, Phys. Rev. \textbf{D57}
  (1998), \href{http://inspirehep.net/search?p=hep-ph/9710416}{2823--2836},
  [\href{http://arXiv.org/pdf/hep-ph/9710416}{{\texttt{ arXiv:hep-ph/9710416}}}
  [hep-ph]]%
\relax\mciteBstWouldAddEndPuncttrue
\mciteSetBstMidEndSepPunct{\mcitedefaultmidpunct}
{\mcitedefaultendpunct}{\mcitedefaultseppunct}\relax
\EndOfBibitem
\bibitem{Grazzini:2013bna}
M.~Grazzini, S.~Kallweit, D.~Rathlev and A.~Torre, \emph{{$Z\gamma$ production
  at hadron colliders in NNLO QCD}}, Phys. Lett. \textbf{B731} (2014),
  \href{http://inspirehep.net/search?p=1309.7000}{204--207},
  [\href{http://arXiv.org/pdf/1309.7000}{{\texttt{ arXiv:1309.7000}}}
  [hep-ph]]%
\relax\mciteBstWouldAddEndPuncttrue
\mciteSetBstMidEndSepPunct{\mcitedefaultmidpunct}
{\mcitedefaultendpunct}{\mcitedefaultseppunct}\relax
\EndOfBibitem
\bibitem{Grazzini:2015nwa}
M.~Grazzini, S.~Kallweit and D.~Rathlev, \emph{{$W\gamma$ and $Z\gamma$
  production at the LHC in NNLO QCD}}, JHEP \textbf{07} (2015),
  \href{http://inspirehep.net/search?p=1504.01330}{085},
  [\href{http://arXiv.org/pdf/1504.01330}{{\texttt{ arXiv:1504.01330}}}
  [hep-ph]]%
\relax\mciteBstWouldAddEndPuncttrue
\mciteSetBstMidEndSepPunct{\mcitedefaultmidpunct}
{\mcitedefaultendpunct}{\mcitedefaultseppunct}\relax
\EndOfBibitem
\bibitem{Campbell:2017aul}
\href{http://inspirehep.net/search?p=1708.02925}{J.~M. Campbell, T.~Neumann and
  C.~Williams}, \emph{{$Z\gamma$ production at NNLO including anomalous
  couplings}},  \href{http://arXiv.org/pdf/1708.02925}{{\texttt{
  arXiv:1708.02925}}} [hep-ph]%
\relax\mciteBstWouldAddEndPuncttrue
\mciteSetBstMidEndSepPunct{\mcitedefaultmidpunct}
{\mcitedefaultendpunct}{\mcitedefaultseppunct}\relax
\EndOfBibitem
\bibitem{Denner:2015fca}
A.~Denner, S.~Dittmaier, M.~Hecht and C.~Pasold, \emph{{NLO QCD and electroweak
  corrections to $Z+\gamma$ production with leptonic Z-boson decays}}, JHEP
  \textbf{02} (2016), \href{http://inspirehep.net/search?p=1510.08742}{057},
  [\href{http://arXiv.org/pdf/1510.08742}{{\texttt{ arXiv:1510.08742}}}
  [hep-ph]]%
\relax\mciteBstWouldAddEndPuncttrue
\mciteSetBstMidEndSepPunct{\mcitedefaultmidpunct}
{\mcitedefaultendpunct}{\mcitedefaultseppunct}\relax
\EndOfBibitem
\bibitem{Catani:2001cc}
S.~Catani, F.~Krauss, R.~Kuhn and B.~R. Webber, \emph{{QCD matrix elements +
  parton showers}}, JHEP \textbf{11} (2001),
  \href{http://inspirehep.net/search?p=hep-ph/0109231}{063},
  [\href{http://arXiv.org/pdf/hep-ph/0109231}{{\texttt{ hep-ph/0109231}}}]%
\relax\mciteBstWouldAddEndPuncttrue
\mciteSetBstMidEndSepPunct{\mcitedefaultmidpunct}
{\mcitedefaultendpunct}{\mcitedefaultseppunct}\relax
\EndOfBibitem
\bibitem{Lonnblad:2001iq}
L.~L{\"o}nnblad, \emph{{Correcting the colour-dipole cascade model with fixed
  order matrix elements}}, JHEP \textbf{05} (2002),
  \href{http://inspirehep.net/search?p=hep-ph/0112284}{046},
  [\href{http://arXiv.org/pdf/hep-ph/0112284}{{\texttt{ hep-ph/0112284}}}]%
\relax\mciteBstWouldAddEndPuncttrue
\mciteSetBstMidEndSepPunct{\mcitedefaultmidpunct}
{\mcitedefaultendpunct}{\mcitedefaultseppunct}\relax
\EndOfBibitem
\bibitem{Krauss:2002up}
F.~Krauss, \emph{{Matrix elements and parton showers in hadronic
  interactions}}, JHEP \textbf{08} (2002),
  \href{http://inspirehep.net/search?p=hep-ph/0205283}{015},
  [\href{http://arXiv.org/pdf/hep-ph/0205283}{{\texttt{ hep-ph/0205283}}}]%
\relax\mciteBstWouldAddEndPuncttrue
\mciteSetBstMidEndSepPunct{\mcitedefaultmidpunct}
{\mcitedefaultendpunct}{\mcitedefaultseppunct}\relax
\EndOfBibitem
\bibitem{Mangano:2001xp}
M.~L. Mangano, M.~Moretti and R.~Pittau, \emph{{Multijet matrix elements and
  shower evolution in hadronic collisions: $W b\bar{b}+n$-jets as a case
  study}}, Nucl. Phys. \textbf{B632} (2002),
  \href{http://inspirehep.net/search?p=hep-ph/0108069}{343--362},
  [\href{http://arXiv.org/pdf/hep-ph/0108069}{{\texttt{ hep-ph/0108069}}}]%
\relax\mciteBstWouldAddEndPuncttrue
\mciteSetBstMidEndSepPunct{\mcitedefaultmidpunct}
{\mcitedefaultendpunct}{\mcitedefaultseppunct}\relax
\EndOfBibitem
\bibitem{Alwall:2007fs}
J.~Alwall et~al., \emph{{Comparative study of various algorithms for the
  merging of parton showers and matrix elements in hadronic collisions}}, Eur.
  Phys. J. \textbf{C53} (2008),
  \href{http://inspirehep.net/record/753397}{473--500},
  [\href{http://arXiv.org/pdf/0706.2569}{{\texttt{ arXiv:0706.2569}}}
  [hep-ph]]%
\relax\mciteBstWouldAddEndPuncttrue
\mciteSetBstMidEndSepPunct{\mcitedefaultmidpunct}
{\mcitedefaultendpunct}{\mcitedefaultseppunct}\relax
\EndOfBibitem
\bibitem{Hamilton:2009ne}
K.~Hamilton, P.~Richardson and J.~Tully, \emph{{A modified CKKW matrix element
  merging approach to angular-ordered parton showers}}, JHEP \textbf{11}
  (2009), \href{http://inspirehep.net/search?p=arXiv:0905.3072}{038},
  [\href{http://arXiv.org/pdf/0905.3072}{{\texttt{ arXiv:0905.3072}}}
  [hep-ph]]%
\relax\mciteBstWouldAddEndPuncttrue
\mciteSetBstMidEndSepPunct{\mcitedefaultmidpunct}
{\mcitedefaultendpunct}{\mcitedefaultseppunct}\relax
\EndOfBibitem
\bibitem{Hoeche:2009rj}
S.~H{\"o}che, F.~Krauss, S.~Schumann and F.~Siegert, \emph{{QCD matrix elements
  and truncated showers}}, JHEP \textbf{05} (2009),
  \href{http://inspirehep.net/search?p=arXiv:0903.1219}{053},
  [\href{http://arXiv.org/pdf/0903.1219}{{\texttt{ arXiv:0903.1219}}}
  [hep-ph]]%
\relax\mciteBstWouldAddEndPuncttrue
\mciteSetBstMidEndSepPunct{\mcitedefaultmidpunct}
{\mcitedefaultendpunct}{\mcitedefaultseppunct}\relax
\EndOfBibitem
\bibitem{Barze:2014zba}
L.~Barze, M.~Chiesa, G.~Montagna, P.~Nason, O.~Nicrosini, F.~Piccinini and
  V.~Prosperi, \emph{{W$\gamma$ production in hadronic collisions using the
  POWHEG+MiNLO method}}, JHEP \textbf{12} (2014),
  \href{http://inspirehep.net/search?p=1408.5766}{039},
  [\href{http://arXiv.org/pdf/1408.5766}{{\texttt{ arXiv:1408.5766}}}
  [hep-ph]]%
\relax\mciteBstWouldAddEndPuncttrue
\mciteSetBstMidEndSepPunct{\mcitedefaultmidpunct}
{\mcitedefaultendpunct}{\mcitedefaultseppunct}\relax
\EndOfBibitem
\bibitem{Hoeche:2012yf}
S.~H{\"o}che, F.~Krauss, M.~Sch{\"o}nherr and F.~Siegert, \emph{{QCD matrix
  elements + parton showers: The NLO case}}, JHEP \textbf{04} (2013),
  \href{http://inspirehep.net/record/1123387}{027},
  [\href{http://arXiv.org/pdf/1207.5030}{{\texttt{ arXiv:1207.5030}}}
  [hep-ph]]%
\relax\mciteBstWouldAddEndPuncttrue
\mciteSetBstMidEndSepPunct{\mcitedefaultmidpunct}
{\mcitedefaultendpunct}{\mcitedefaultseppunct}\relax
\EndOfBibitem
\bibitem{Hoeche:2011fd}
S.~H{\"o}che, F.~Krauss, M.~Sch{\"o}nherr and F.~Siegert, \emph{{A critical
  appraisal of NLO+PS matching methods}}, JHEP \textbf{09} (2012),
  \href{http://inspirehep.net/record/944643}{049},
  [\href{http://arXiv.org/pdf/1111.1220}{{\texttt{ arXiv:1111.1220}}}
  [hep-ph]]%
\relax\mciteBstWouldAddEndPuncttrue
\mciteSetBstMidEndSepPunct{\mcitedefaultmidpunct}
{\mcitedefaultendpunct}{\mcitedefaultseppunct}\relax
\EndOfBibitem
\bibitem{Hoeche:2012ft}
S.~H{\"o}che, F.~Krauss, M.~Sch{\"o}nherr and F.~Siegert, \emph{{W+n-jet
  predictions with MC@NLO in Sherpa}}, Phys.Rev.Lett. \textbf{110} (2013),
  \href{http://inspirehep.net/record/1086175}{052001},
  [\href{http://arXiv.org/pdf/1201.5882}{{\texttt{ arXiv:1201.5882}}}
  [hep-ph]]%
\relax\mciteBstWouldAddEndPuncttrue
\mciteSetBstMidEndSepPunct{\mcitedefaultmidpunct}
{\mcitedefaultendpunct}{\mcitedefaultseppunct}\relax
\EndOfBibitem
\bibitem{Frixione:2002ik}
S.~Frixione and B.~R. Webber, \emph{{Matching NLO QCD computations and parton
  shower simulations}}, JHEP \textbf{06} (2002),
  \href{http://inspirehep.net/search?p=hep-ph/0204244}{029},
  [\href{http://arXiv.org/pdf/hep-ph/0204244}{{\texttt{ hep-ph/0204244}}}]%
\relax\mciteBstWouldAddEndPuncttrue
\mciteSetBstMidEndSepPunct{\mcitedefaultmidpunct}
{\mcitedefaultendpunct}{\mcitedefaultseppunct}\relax
\EndOfBibitem
\bibitem{Yennie:1961ad}
D.~R. Yennie, S.~C. Frautschi and H.~Suura, \emph{{The Infrared Divergence
  Phenomena and High-Energy Processes}}, Ann. Phys. \textbf{13} (1961),
  \href{http://inspirehep.net/search?j=APNYA,13,379}{379--452}%
\relax\mciteBstWouldAddEndPuncttrue
\mciteSetBstMidEndSepPunct{\mcitedefaultmidpunct}
{\mcitedefaultendpunct}{\mcitedefaultseppunct}\relax
\EndOfBibitem
\bibitem{Schonherr:2008av}
M.~Sch\"{o}nherr and F.~Krauss, \emph{Soft photon radiation in particle decays
  in \Sherpa{}}, JHEP \textbf{12} (2008),
  \href{http://inspirehep.net/search?p=arXiv:0810.5071}{018},
  [\href{http://arXiv.org/pdf/0810.5071}{{\texttt{ arXiv:0810.5071}}}
  [hep-ph]]%
\relax\mciteBstWouldAddEndPuncttrue
\mciteSetBstMidEndSepPunct{\mcitedefaultmidpunct}
{\mcitedefaultendpunct}{\mcitedefaultseppunct}\relax
\EndOfBibitem
\bibitem{Frixione:1998jh}
S.~Frixione, \emph{{Isolated photons in perturbative QCD}}, Phys. Lett.
  \textbf{B429} (1998),
  \href{http://inspirehep.net/search?p=hep-ph/9801442}{369--374},
  [\href{http://arXiv.org/pdf/hep-ph/9801442}{{\texttt{ hep-ph/9801442}}}]%
\relax\mciteBstWouldAddEndPuncttrue
\mciteSetBstMidEndSepPunct{\mcitedefaultmidpunct}
{\mcitedefaultendpunct}{\mcitedefaultseppunct}\relax
\EndOfBibitem
\bibitem{Gleisberg:2008ta}
T.~Gleisberg, S.~H{\"o}che, F.~Krauss, M.~Sch\"{o}nherr, S.~Schumann,
  F.~Siegert and J.~Winter, \emph{{Event generation with \Sherpa 1.1}}, JHEP
  \textbf{02} (2009), \href{http://inspirebeta.net/record/803708}{007},
  [\href{http://arXiv.org/pdf/0811.4622}{{\texttt{ arXiv:0811.4622}}}
  [hep-ph]]%
\relax\mciteBstWouldAddEndPuncttrue
\mciteSetBstMidEndSepPunct{\mcitedefaultmidpunct}
{\mcitedefaultendpunct}{\mcitedefaultseppunct}\relax
\EndOfBibitem
\bibitem{Krauss:2001iv}
F.~Krauss, R.~Kuhn and G.~Soff, \emph{{AMEGIC++ 1.0: A Matrix Element Generator
  In C++}}, JHEP \textbf{02} (2002),
  \href{http://inspirehep.net/search?p=hep-ph/0109036}{044},
  [\href{http://arXiv.org/pdf/hep-ph/0109036}{{\texttt{ hep-ph/0109036}}}]%
\relax\mciteBstWouldAddEndPuncttrue
\mciteSetBstMidEndSepPunct{\mcitedefaultmidpunct}
{\mcitedefaultendpunct}{\mcitedefaultseppunct}\relax
\EndOfBibitem
\bibitem{Gleisberg:2008fv}
T.~Gleisberg and S.~H{\"o}che, \emph{{Comix, a new matrix element generator}},
  JHEP \textbf{12} (2008), \href{http://inspirehep.net/record/793879}{039},
  [\href{http://arXiv.org/pdf/0808.3674}{{\texttt{ arXiv:0808.3674}}}
  [hep-ph]]%
\relax\mciteBstWouldAddEndPuncttrue
\mciteSetBstMidEndSepPunct{\mcitedefaultmidpunct}
{\mcitedefaultendpunct}{\mcitedefaultseppunct}\relax
\EndOfBibitem
\bibitem{Cascioli:2011va}
F.~Cascioli, P.~Maierh{\"o}fer and S.~Pozzorini, \emph{{Scattering Amplitudes
  with Open Loops}}, Phys.Rev.Lett. \textbf{108} (2012),
  \href{http://inspirehep.net/record/946998}{111601},
  [\href{http://arXiv.org/pdf/1111.5206}{{\texttt{ arXiv:1111.5206}}}
  [hep-ph]]%
\relax\mciteBstWouldAddEndPuncttrue
\mciteSetBstMidEndSepPunct{\mcitedefaultmidpunct}
{\mcitedefaultendpunct}{\mcitedefaultseppunct}\relax
\EndOfBibitem
\bibitem{Ossola:2007ax}
G.~Ossola, C.~G. Papadopoulos and R.~Pittau, \emph{{CutTools: A Program
  implementing the OPP reduction method to compute one-loop amplitudes}}, JHEP
  \textbf{0803} (2008), \href{http://inspirehep.net/search?p=0711.3596}{042},
  [\href{http://arXiv.org/pdf/0711.3596}{{\texttt{ arXiv:0711.3596}}}
  [hep-ph]]%
\relax\mciteBstWouldAddEndPuncttrue
\mciteSetBstMidEndSepPunct{\mcitedefaultmidpunct}
{\mcitedefaultendpunct}{\mcitedefaultseppunct}\relax
\EndOfBibitem
\bibitem{vanHameren:2010cp}
A.~van Hameren, \emph{{OneLOop: For the evaluation of one-loop scalar
  functions}}, Comput.Phys.Commun. \textbf{182} (2011),
  \href{http://inspirehep.net/search?p=1007.4716}{2427--2438},
  [\href{http://arXiv.org/pdf/1007.4716}{{\texttt{ arXiv:1007.4716}}}
  [hep-ph]]%
\relax\mciteBstWouldAddEndPuncttrue
\mciteSetBstMidEndSepPunct{\mcitedefaultmidpunct}
{\mcitedefaultendpunct}{\mcitedefaultseppunct}\relax
\EndOfBibitem
\bibitem{Butterworth:2014efa}
\href{http://inspirehep.net/search?p=1405.1067}{J.~Butterworth, G.~Dissertori,
  S.~Dittmaier, D.~de~Florian, N.~Glover et~al.}, \emph{{Les Houches 2013:
  Physics at TeV Colliders: Standard Model Working Group Report}},
  \href{http://arXiv.org/pdf/1405.1067}{{\texttt{ arXiv:1405.1067}}} [hep-ph]%
\relax\mciteBstWouldAddEndPuncttrue
\mciteSetBstMidEndSepPunct{\mcitedefaultmidpunct}
{\mcitedefaultendpunct}{\mcitedefaultseppunct}\relax
\EndOfBibitem
\bibitem{DENNER200622}
A.~Denner and S.~Dittmaier, \emph{{The complex-mass scheme for perturbative
  calculations with unstable particles}}, Nuclear Physics B - Proceedings
  Supplements \textbf{160} (2006),
  \href{http://www.sciencedirect.com/science/article/pii/S0920563206006177}{22
  -- 26}%
\relax\mciteBstWouldAddEndPuncttrue
\mciteSetBstMidEndSepPunct{\mcitedefaultmidpunct}
{\mcitedefaultendpunct}{\mcitedefaultseppunct}\relax
\EndOfBibitem
\bibitem{Ball:2014uwa}
R.~D. Ball et~al., NNPDF, \emph{{Parton distributions for the LHC Run II}},
  JHEP \textbf{04} (2015),
  \href{http://inspirehep.net/search?p=1410.8849}{040},
  [\href{http://arXiv.org/pdf/1410.8849}{{\texttt{ arXiv:1410.8849}}}
  [hep-ph]]%
\relax\mciteBstWouldAddEndPuncttrue
\mciteSetBstMidEndSepPunct{\mcitedefaultmidpunct}
{\mcitedefaultendpunct}{\mcitedefaultseppunct}\relax
\EndOfBibitem
\bibitem{Sjostrand:1987su}
T.~Sj{\"o}strand and M.~van Zijl, \emph{{A multiple-interaction model for the
  event structure in hadron collisions}}, Phys. Rev. \textbf{D36} (1987),
  \href{http://inspirehep.net/search?j=PHRVA,D36,2019}{2019}%
\relax\mciteBstWouldAddEndPuncttrue
\mciteSetBstMidEndSepPunct{\mcitedefaultmidpunct}
{\mcitedefaultendpunct}{\mcitedefaultseppunct}\relax
\EndOfBibitem
\bibitem{Alekhin:2005dx}
\href{http://inspirehep.net/search?p=hep-ph/0601012}{S.~Alekhin et~al.},
  \emph{{HERA and the LHC - A workshop on the implications of HERA for LHC
  physics: Proceedings Part A}},
  \href{http://arXiv.org/pdf/hep-ph/0601012}{{\texttt{ hep-ph/0601012}}}%
\relax\mciteBstWouldAddEndPuncttrue
\mciteSetBstMidEndSepPunct{\mcitedefaultmidpunct}
{\mcitedefaultendpunct}{\mcitedefaultseppunct}\relax
\EndOfBibitem
\bibitem{Winter:2003tt}
J.-C. Winter, F.~Krauss and G.~Soff, \emph{{A modified cluster-hadronisation
  model}}, Eur. Phys. J. \textbf{C36} (2004),
  \href{http://inspirehep.net/search?p=hep-ph/0311085}{381--395},
  [\href{http://arXiv.org/pdf/hep-ph/0311085}{{\texttt{ hep-ph/0311085}}}]%
\relax\mciteBstWouldAddEndPuncttrue
\mciteSetBstMidEndSepPunct{\mcitedefaultmidpunct}
{\mcitedefaultendpunct}{\mcitedefaultseppunct}\relax
\EndOfBibitem
\bibitem{Buckley:2010ar}
A.~Buckley, J.~Butterworth, L.~L{\"o}nnblad, D.~Grellscheid, H.~Hoeth et~al.,
  \emph{{Rivet user manual}}, Comput.Phys.Commun. \textbf{184} (2013),
  \href{http://inspirehep.net/search?p=1003.0694}{2803--2819},
  [\href{http://arXiv.org/pdf/1003.0694}{{\texttt{ arXiv:1003.0694}}}
  [hep-ph]]%
\relax\mciteBstWouldAddEndPuncttrue
\mciteSetBstMidEndSepPunct{\mcitedefaultmidpunct}
{\mcitedefaultendpunct}{\mcitedefaultseppunct}\relax
\EndOfBibitem
\bibitem{Catani:1993hr}
S.~Catani, Y.~L. Dokshitzer, M.~H. Seymour and B.~R. Webber,
  \emph{{Longitudinally-invariant $k_\perp$-clustering algorithms for
  hadron--hadron collisions}}, Nucl. Phys. \textbf{B406} (1993),
  \href{http://inspirehep.net/search?j=NUPHA,B406,187}{187--224}%
\relax\mciteBstWouldAddEndPuncttrue
\mciteSetBstMidEndSepPunct{\mcitedefaultmidpunct}
{\mcitedefaultendpunct}{\mcitedefaultseppunct}\relax
\EndOfBibitem
\bibitem{Dixon:1999abc}
L.~Dixon, Z.~Kunszt and A.~Signer, \emph{Vector boson pair production in
  hadronic collisions at $O({\ensuremath{\alpha}}_{s}):$ Lepton correlations
  and anomalous couplings}, Phys. Rev. D \textbf{60} (1999),
  \href{https://link.aps.org/doi/10.1103/PhysRevD.60.114037}{114037}%
\relax\mciteBstWouldAddEndPuncttrue
\mciteSetBstMidEndSepPunct{\mcitedefaultmidpunct}
{\mcitedefaultendpunct}{\mcitedefaultseppunct}\relax
\EndOfBibitem
\bibitem{Denner201518}
A.~Denner, S.~Dittmaier, M.~Hecht and C.~Pasold, \emph{{NLO QCD and electroweak
  corrections to $W \gamma$ production with leptonic $W$-boson decays}},
  Journal of High Energy Physics \textbf{2015} (2015), no.~4,
  \href{http://dx.doi.org/10.1007/JHEP04(2015)018}{18}%
\relax\mciteBstWouldAddEndPuncttrue
\mciteSetBstMidEndSepPunct{\mcitedefaultmidpunct}
{\mcitedefaultendpunct}{\mcitedefaultseppunct}\relax
\EndOfBibitem
\bibitem{PhysRevD.93.112002}
G.~Aad et~al., ATLAS Collaboration, \emph{{Measurements of
  $Z\ensuremath{\gamma}$ and $Z\ensuremath{\gamma\gamma}$ production in pp
  collisions at $\sqrt{s}=8$ TeV with the ATLAS detector}}, Physical Review D
  \textbf{93} (2016), no.~11%
\relax\mciteBstWouldAddEndPunctfalse
\mciteSetBstMidEndSepPunct{\mcitedefaultmidpunct}
{}{\mcitedefaultseppunct}\relax
\EndOfBibitem
\bibitem{Hoeche:2009xc}
S.~H{\"o}che, S.~Schumann and F.~Siegert, \emph{{Hard photon production and
  matrix-element parton-shower merging}}, Phys. Rev. \textbf{D81} (2010),
  \href{http://inspirehep.net/search?p=arXiv:0912.3501}{034026},
  [\href{http://arXiv.org/pdf/0912.3501}{{\texttt{ arXiv:0912.3501}}}
  [hep-ph]]%
\relax\mciteBstWouldAddEndPuncttrue
\mciteSetBstMidEndSepPunct{\mcitedefaultmidpunct}
{\mcitedefaultendpunct}{\mcitedefaultseppunct}\relax
\EndOfBibitem
\end{mcitethebibliography}

\end{document}